\documentclass[aps,prb,twocolumn,superscriptaddress,longbibliography]{revtex4-2}
\usepackage[T1]{fontenc}
\usepackage[english]{babel}
\usepackage[inline]{enumitem}
\usepackage{glossaries}
\usepackage{graphicx}
\usepackage{xcolor}
\usepackage{amsmath}
\usepackage{amssymb}
\usepackage{multirow}
\usepackage{booktabs}
\usepackage{hyperref}
\usepackage{etoolbox}
\usepackage{textcomp}

\definecolor{DARKMAGENTA}{HTML}{AF2F40}
\hypersetup{
    colorlinks=true,
    linkcolor=DARKMAGENTA,
    citecolor=DARKMAGENTA,
    urlcolor=DARKMAGENTA,
}

\usepackage{listings}
\newacronym{dft}{DFT}{density-functional theory}
\newacronym{eos}{EOS}{equation of state}
\newacronym{dc}{DC}{dissociation curve}
\newacronym{ase}{ASE}{Atomistic Simulation Environment}
\newacronym{asr}{ASR}{Atomistic Simulation Recipes}
\newacronym{hpc}{HPC}{high-performance computing}
\newacronym{api}{API}{application programming interface}
\newacronym{gui}{GUI}{graphical user interface}
\newacronym{pbe}{PBE}{Perdew–Burke-Ernzerhof}
\newacronym{bcc}{BCC}{body-centered cubic}

\newcommand\cfref[1]{Fig.~\ref{#1}}
\newcommand\csref[1]{Sec.~\ref{#1}}
\newcommand\ctref[1]{Tab.~\ref{#1}}
\newcommand\clref[1]{Listing~\ref{#1}}

\newcommand\qe{\textsc{Quantum ESPRESSO}}
\newcommand\siesta{\textsc{Siesta}}
\newcommand\cptwok{CP2K}
\newcommand\fleur{FLEUR}
\newcommand\abinit{\textsc{Abinit}}
\newcommand\vasp{VASP}
\newcommand\bigdft{BigDFT}
\newcommand\castep{CASTEP}
\newcommand\nwchem{NWChem}
\newcommand\gaussian{Gaussian}
\newcommand\orca{ORCA}

\usepackage{pifont}
\usepackage{tabularx}

\definecolor{RED}{HTML}{AF2F40}
\definecolor{ORANGE}{HTML}{E38851}
\definecolor{GREEN}{HTML}{40B46F}

\newacronym{cli}{CLI}{command line interface}
\newacronym{sssp}{SSSP}{Standard Solid State Pseudopotentials}

\newcommand{\cmark}{{\textcolor{GREEN}{\ding{51}}}}
\newcommand{\xmark}{{\textcolor{RED}{\ding{54}}}}
\newcommand{\nmark}{{\textcolor{ORANGE}{\ding{108}}}}
\newcommand\bohr{bohr}

\begin{document}

\title{Common workflows for computing material properties using different quantum engines}

\author{Sebastiaan P. Huber}
\email{mail@sphuber.net}
\affiliation{Theory and Simulation of Materials (THEOS) and National Centre for Computational Design and Discovery of Novel Materials (MARVEL), \'Ecole Polytechnique F\'ed\'erale de Lausanne, CH-1015 Lausanne, Switzerland}

\author{Emanuele Bosoni}
\affiliation{Institut de Ci\`encia de Materials de Barcelona, ICMAB-CSIC, Campus UAB, 08193 Bellaterra, Spain}

\author{Marnik Bercx}
\affiliation{Theory and Simulation of Materials (THEOS) and National Centre for Computational Design and Discovery of Novel Materials (MARVEL), \'Ecole Polytechnique F\'ed\'erale de Lausanne, CH-1015 Lausanne, Switzerland}

\author{Jens Br\"oder}
\affiliation{Peter Gr\"unberg Institut and Institute for Advanced Simulation, Forschungszentrum J\"ulich, D-52425 J\"ulich, Germany}
\affiliation{Department of Physics, RWTH Aachen University, D-52056, Aachen, Germany}

\author{Augustin Degomme}
\affiliation{Univ. Grenoble-Alpes, CEA, IRIG-MEM-L\_Sim, 38000 Grenoble, France}

\author{Vladimir Dikan}
\affiliation{Institut de Ci\`encia de Materials de Barcelona, ICMAB-CSIC, Campus UAB, 08193 Bellaterra, Spain}

\author{Kristjan Eimre}
\affiliation{nanotech@surfaces laboratory, Swiss Federal Laboratories for Materials Science and Technology (Empa), CH-8600 D\"ubendorf, Switzerland}

\author{Espen Flage-Larsen}
\affiliation{SINTEF Industry, Materials Physics, Oslo, Norway}
\affiliation{University of Oslo, Department of Physics, Norway}

\author{Alberto Garcia}
\affiliation{Institut de Ci\`encia de Materials de Barcelona, ICMAB-CSIC, Campus UAB, 08193 Bellaterra, Spain}

\author{Luigi Genovese}
\affiliation{Univ. Grenoble-Alpes, CEA, IRIG-MEM-L\_Sim, 38000 Grenoble, France}

\author{Dominik Gresch}
\affiliation{Microsoft Station Q, University of California, Santa Barbara, California, 93106-6105, USA}

\author{Conrad Johnston}
\affiliation{Atomistic Simulation Centre, School of Mathematics and Physics, Queen's University Belfast, United Kingdom}

\author{Guido Petretto}
\affiliation{UCLouvain, Institut de la Mati\`ere Condens\'ee et des Nanosciences (IMCN), Chemin des \'Etoiles~8, Louvain-la-Neuve 1348, Belgium}

\author{Samuel Ponc\'e}
\affiliation{Theory and Simulation of Materials (THEOS) and National Centre for Computational Design and Discovery of Novel Materials (MARVEL), \'Ecole Polytechnique F\'ed\'erale de Lausanne, CH-1015 Lausanne, Switzerland}

\author{Gian-Marco Rignanese}
\affiliation{UCLouvain, Institut de la Mati\`ere Condens\'ee et des Nanosciences (IMCN), Chemin des \'Etoiles~8, Louvain-la-Neuve 1348, Belgium}

\author{Christopher J. Sewell}
\affiliation{Theory and Simulation of Materials (THEOS) and National Centre for Computational Design and Discovery of Novel Materials (MARVEL), \'Ecole Polytechnique F\'ed\'erale de Lausanne, CH-1015 Lausanne, Switzerland}

\author{Berend Smit}
\affiliation{Laboratory of Molecular Simulation (LSMO), Institut des sciences et ing\'enierie chimiques (ISIC), \'Ecole Polytechnique F\'ed\'erale de Lausanne (EPFL) Valais, CH-1951, Sion, Switzerland}

\author{Vasily Tseplyaev}
\affiliation{Peter Gr\"unberg Institut and Institute for Advanced Simulation, Forschungszentrum J\"ulich, D-52425 J\"ulich, Germany}
\affiliation{Department of Physics, RWTH Aachen University, D-52056, Aachen, Germany}

\author{Martin Uhrin}
\affiliation{Theory and Simulation of Materials (THEOS) and National Centre for Computational Design and Discovery of Novel Materials (MARVEL), \'Ecole Polytechnique F\'ed\'erale de Lausanne, CH-1015 Lausanne, Switzerland}

\author{Daniel Wortmann}
\affiliation{Peter Gr\"unberg Institut and Institute for Advanced Simulation, Forschungszentrum J\"ulich, D-52425 J\"ulich, Germany}

\author{Aliaksandr V. Yakutovich}
\affiliation{Laboratory of Molecular Simulation (LSMO), Institut des sciences et ing\'enierie chimiques (ISIC), \'Ecole Polytechnique F\'ed\'erale de Lausanne (EPFL) Valais, CH-1951, Sion, Switzerland}
\affiliation{Theory and Simulation of Materials (THEOS) and National Centre for Computational Design and Discovery of Novel Materials (MARVEL), \'Ecole Polytechnique F\'ed\'erale de Lausanne, CH-1015 Lausanne, Switzerland}

\author{Austin Zadoks}
\affiliation{Theory and Simulation of Materials (THEOS) and National Centre for Computational Design and Discovery of Novel Materials (MARVEL), \'Ecole Polytechnique F\'ed\'erale de Lausanne, CH-1015 Lausanne, Switzerland}

\author{Pezhman Zarabadi-Poor}
\affiliation{Department of Chemistry, Claverton Down, University of Bath, BA2 7AY, Bath, United Kingdom}
\affiliation{The Faraday Institution, Didcot OX11 0RA, United Kingdom}

\author{Bonan Zhu}
\affiliation{Department of Chemistry, University College London, 20 Gordon St, Bloomsbury, London WC1H 0AJ, United Kingdom}
\affiliation{The Faraday Institution, Didcot OX11 0RA, United Kingdom}

\author{Nicola Marzari}
\affiliation{Theory and Simulation of Materials (THEOS) and National Centre for Computational Design and Discovery of Novel Materials (MARVEL), \'Ecole Polytechnique F\'ed\'erale de Lausanne, CH-1015 Lausanne, Switzerland}

\author{Giovanni Pizzi}
\email{giovanni.pizzi@epfl.ch}
\affiliation{Theory and Simulation of Materials (THEOS) and National Centre for Computational Design and Discovery of Novel Materials (MARVEL), \'Ecole Polytechnique F\'ed\'erale de Lausanne, CH-1015 Lausanne, Switzerland}

\date{\today}

\begin{abstract}
The prediction of material properties through electronic-structure simulations based on density-functional theory has become routinely common, thanks, in part, to the steady increase in the number and robustness of available simulation packages.
This plurality of codes and methods aiming to solve similar problems is both a boon and a burden.
While providing great opportunities for cross-verification, these packages adopt different methods, algorithms, and paradigms, making it challenging to choose, master, and efficiently use any one for a given task.
Leveraging recent advances in managing reproducible scientific workflows, we demonstrate how developing common interfaces for workflows that automatically compute material properties can tackle the challenge mentioned above, greatly simplifying interoperability and cross-verification.
We introduce design rules for reproducible and reusable code-agnostic workflow interfaces to compute well-defined material properties, which we implement for eleven different quantum engines and use to compute three different material properties.
Each implementation encodes carefully selected simulation parameters and workflow logic, making the implementer's expertise of the quantum engine directly available to non-experts.
Full provenance and reproducibility of the workflows is guaranteed through the use of the AiiDA infrastructure.
All workflows are made available as open-source and come pre-installed with the Quantum Mobile virtual machine, making their use straightforward.
\end{abstract}


\maketitle

\section{Introduction}
The use of \gls{dft} to compute the properties of systems at the atomic level has become widespread~\cite{Burke:2012,Jones:2015}, as both the number of quantum engines that implement it and the available computational power continue to increase.
However, despite its large-scale deployment both in academia and in industry, the application of \gls{dft} is still not a trivial operation.
Accurate predictions require expert knowledge of not just \gls{dft} itself but also of the specific code used to perform the calculations (throughout this work we will use the terms quantum engine and code interchangeably).
Although the diversity of available simulation packages improves the accuracy and reliability of results by virtue of cross-verification\cite{Lejaeghere:2016}, different codes use diverse computational methods and interfaces, making it difficult even for experts to master more than just a few of them.
This may result in software being used not for its applicability to a particular problem, but merely due to circumstantial reasons.
Furthermore, the fact that the correct usage of \gls{dft}-based codes requires expert knowledge directly limits its application and potential for scientific discovery.

Although \gls{dft} is used to compute many material properties of varying complexity, a large percentage of all performed calculations are defined by relatively simple recipes.
Therefore, in addition to implementing new functionalities and improving the accuracy of existing ones, the effort of domain and code experts should be focused on providing robust workflows with common interfaces that can be used by experts and non-experts alike.
If these are designed properly such that they are reusable, they can be employed as modular blocks in building more complex workflows, e.g. in a multi-scale approach.
On top of reusability, in order to guarantee that results can be validated, it is crucial that these common workflows are reproducible.

The \gls{ase}~\cite{HjorthLarsen:2017} initiated an effort to provide a single interface for various quantum engines, which was later extended with the \gls{asr}~\cite{Gjerding:2021}.
In this article, we address the additional challenges that one faces when trying to develop a common workflow interface, focusing particularly on the requirements of reusability and reproducibility, and we provide a solution based on AiiDA, an informatics infrastructure and workflow management system~\cite{Huber:2020}.
As a proof-of-concept, we define a common workflow interface specifically for the optimization of solid-state structures and molecular geometries, together with its implementation in eleven quantum codes: \abinit{}\cite{Gonze:2016,Gonze:2020,Romero:2020}, \bigdft{}\cite{Ratcliff:2020}, \castep{}\cite{Clark:2005}, \cptwok{}\cite{Hutter:2013,Kuehne:2020}, \fleur{}\cite{fleur}, \gaussian{}\cite{Frisch:2016}, \nwchem{}\cite{Apra:2020}, \orca{}\cite{Neese:2012,Neese:2018}, \qe{}\cite{Giannozzi:2009,Giannozzi:2017}, \siesta{}\cite{Soler:2002,Garcia:2020} and \vasp{}\cite{Kresse:1996,Kresse:1999}.
This particular common workflow interface, referred to as the ``common relax workflow'' throughout this work, allows a user to optimize a structure using any of these codes without having to define code specific parameters.
The computed results are returned in a single unified format with identical units making the results directly comparable and reusable regardless of the underlying quantum engine used.

Each implementation of the common relax workflow interface provides at least three protocols (`fast', `moderate' and `precise') that allow a user to specify the desired computational accuracy in an intuitive and general way.
The mapping between these levels of protocols and code specific parameters are up to the respective code experts to define.
Through these protocols, expert knowledge of appropriate numerical parameters is thus encoded directly into the workflows, reducing the risk of incorrect or unreliable results and opening up the use of the quantum engines also to non-experts.
Despite the ease-of-use of the workflows, the workflow interface design (which will be discussed later) maintains full flexibility and allows users to override any of the sensible defaults provided by the protocols.
Furthermore, since AiiDA tracks the full provenance graph of executed workflows, storing all parameters used in workflow steps, the appropriateness of the inputs and the correctness of the results can also be checked \emph{a posteriori}.

To demonstrate the concept of modularity and potential for cross-verification, we use the common relax workflow to compute the \gls{eos} and the \gls{dc}, which are commonly computed properties for bulk compounds and diatomic molecules, respectively.
Each of these properties is computed by a single workflow that exclusively leverages the common relax workflow as a modular building block, allowing any of the quantum engines to be used without specifying any code-specific parameters.
The \gls{eos} and the \gls{dc} are computed for a few compounds with different geometric, electronic and magnetic properties.
As we will show later, the results computed by the various quantum engines show good agreement.
We stress here that the focus of this paper is not on the validation of the results, but rather on the demonstration of the feasibility of a common workflow interface, directly enabling the reusability of complex workflows and the cross-verification of their results.
We hope this will motivate readers to generalize these concepts and apply them to a broader and more complex range of problems.

The implementations of the common relax workflow interface of all quantum engines described in this paper are made available as free open-source software at \href{https://github.com/aiidateam/aiida-common-workflows}{https://github.com/aiidateam/aiida-common-workflows} under the MIT license.
In addition, all workflows, as well as the seven quantum engines with a free open-source software license, come pre-installed in the Quantum Mobile\cite{Talirz:2020} virtual machine (and quantum engines with a more restrictive license can be manually installed on any computational resource and configured to be used with AiiDA).
This makes it straightforward to fully reproduce all the results presented in this paper (see the Supplementary Information for complete instructions).

\section{Results and Discussion}
\subsection{Reusability and reproducibility}
\label{sec:reusability}
Workflows, by definition, consist of multiple steps or multiple subprocesses that are executed in series, in parallel, or in a combination thereof, to obtain the final result.
Ideally, workflows can themselves be used as modular blocks, becoming steps of higher-level workflows.
To keep this process practical and tractable, workflows should be designed to be as modular and reusable as possible.
Additionally, as workflows become ever more complex, so does their reproducibility.
In this paper, we focus on two particular concepts that address these requirements: optional transparency and scoped provenance.

\subsubsection{Optional transparency}
In software design, the term \emph{transparency} is often used to mean that a consumer of an interface should not be bothered with the inner details of the implementation (the details are invisible, or transparent).
In terms of computational workflows, this can be taken to mean that a useful generic turn-key solution should have a simple interface, requiring as few inputs as possible from the user.
Apart from physical inputs (e.g., the initial crystal structure in a relaxation workflow) and flags to determine which type of simulation to run (e.g., relax only atomic positions or also the periodic cell), any other input that is only needed as a numerical parameter by the underlying implementation should be automatically determined by the workflow.

\begin{figure*}[ht!]
    \centering
    \includegraphics[width=0.9\linewidth]{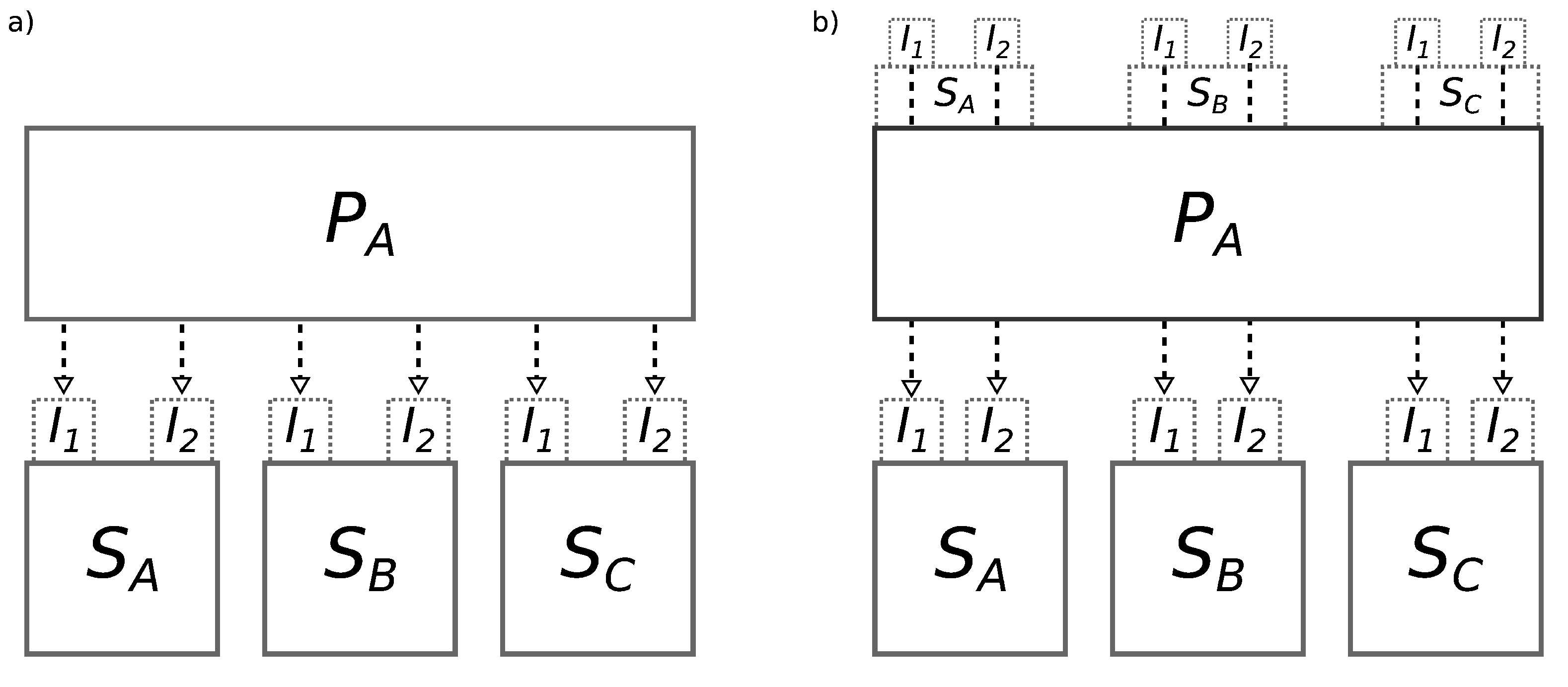}
    \caption{{\bfseries Difference between a transparent and opaque workflow interface.}
        a) A schematic depiction of a process ($P_A$) that consists of three subprocesses ($S_A$, $S_B$ and $S_C$), that each require two inputs ($I_1$ and $I_2$).
        In this abstract example, the top-level process takes no inputs which is just for clarity; normally the top-level process takes at least one input based on which the inputs for the subprocesses are determined.
        Note that, although for simplicity the same symbol is used for these inputs, they do not necessarily represent identical inputs across the subprocesses, even though in practice the names could actually overlap.
        Only two inputs per process are arbitrarily chosen here for illustrative purposes.
        The interface of $P_A$ does not expose the inputs of its subprocesses, but instead will decide them internally.
        This means that a user of $P_A$ cannot customise the inputs of any of subprocesses.
        b) A schematic depiction of the same process $P_A$ as in a), but in this case exposing the inputs of its subprocesses.
        Since the names of the inputs can potentially overlap, inputs are exposed in namespaces to prevent name clashes.
        A user of $P_A$ can now directly set the inputs through the top-level interface.
        If any of the inputs of the subprocesses \emph{should not} be defined by the user (due to being part of the workflow's task to define it) the workflow can decide to not expose that particular input.
    }
    \label{fig:schematic-process-interface}
\end{figure*}

However, this transparency of the interface comes at a cost.
Complex workflows often consist of multiple subprocesses, each requiring their own inputs.
Oftentimes at least some of these inputs cannot be automatically determined by the main workflow, as they are circumstantial and will be dependent on \emph{how} and \emph{where} the workflow is run.
An example is when one of the subprocesses is executed on a \gls{hpc} cluster and therefore requires specific environmental settings, such as the required resources and parallelization flags.
A transparent interface is closed to these inputs being set (as shown schematically in \cfref{fig:schematic-process-interface}a) and, as such, the workflow will be tied to a very specific environment for execution.
Therefore, it will not be portable and consequently not reusable.
But even if the inputs of the workflow could be automatically determined, an expert user may still want to override them.
Transparent interfaces precluding this level of control diminish the reusability of workflows.

The solution to the aforementioned problem is to make the interface for all workflows fully \emph{opaque} and expose all inputs of their subprocesses.
That is to say, the workflow should make it possible to define each and every input that any of its subprocesses takes, as shown in \cfref{fig:schematic-process-interface}b.
By doing so, a user has access to all the inputs of the subprocesses, whether they could have been automatically determined by the workflow or not.
Certainly, there are situations where the workflow can consciously decide not to expose certain inputs, as it is part of its task to determine them based on other inputs or intermediate results.

We are now confronted with two conflicting requirements, where a workflow interface must be both \emph{transparent} for simplicity, yet at the same time fully \emph{opaque} for reusability.
The solution is to create an interface that is \emph{optionally transparent}, i.e., it is opaque when needed but can still be used in a transparent manner whenever possible.
Exposing the inputs of subprocesses is the first crucial step towards obtaining this goal, but it is not the only one.
In addition, the workflow needs to specify sensible defaults such that the interface remains simple to operate with just a minimal set of inputs.
An even better solution is offered by what we refer to as \emph{input generators}.
An input generator for a workflow is a function that, based on a minimal set of essential inputs, generates the full set of inputs required by the workflow and all of its subprocesses.
The advantage of this approach is that an expert user has the ability to inspect the full set of inputs that have been generated and even modify them before actually executing the workflow.
This is the approach that we will take in the rest of this work.

\subsubsection{Scoped provenance}
\begin{figure*}[ht!]
    \centering
    \includegraphics[width=0.9\linewidth]{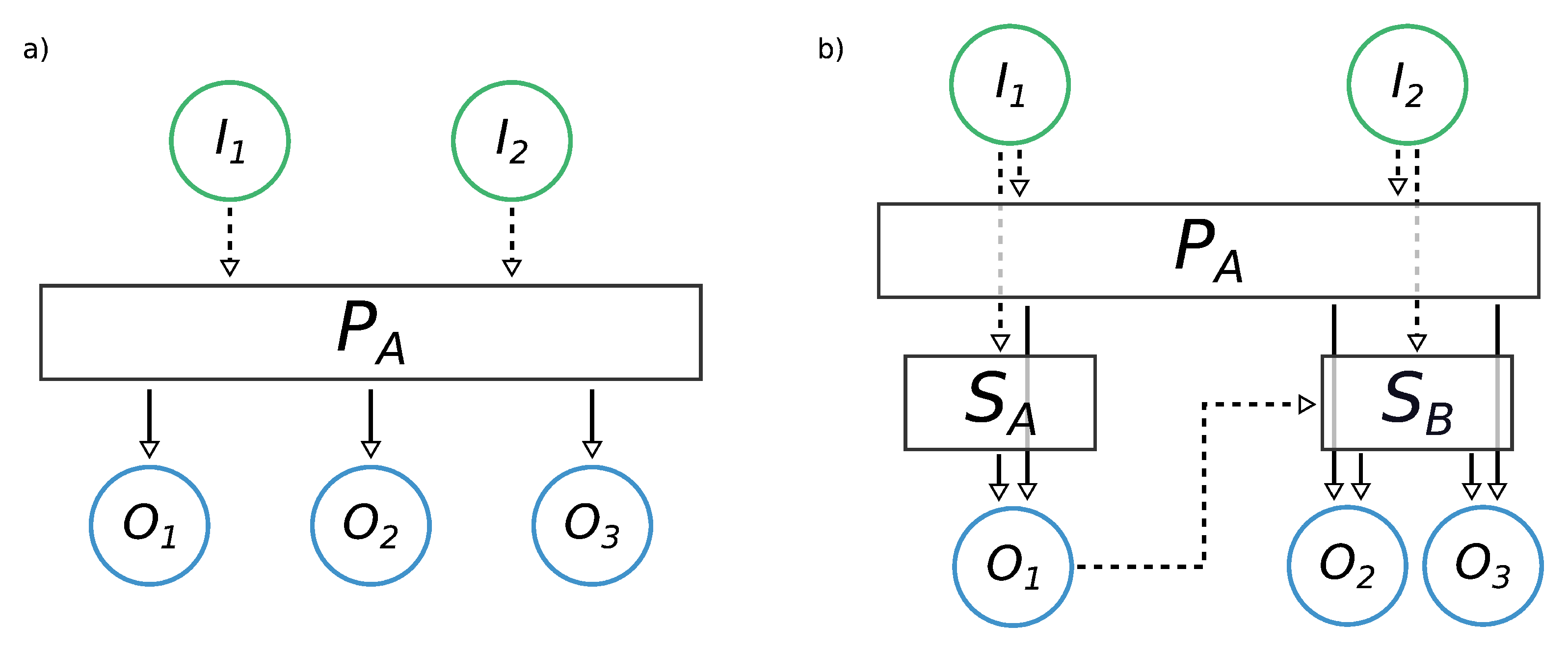}
    \caption{{\bfseries Scope provenance.}
        a) Schematic provenance of a workflow ($P_A$) that takes two inputs ($I_1$ and $I_2$) and produces three outputs ($O_1$, $O_2$, $O_3$).
        b) A more detailed view of the complete provenance of $P_A$, which actually runs two subprocesses ($S_A$ and $S_B$).
        Input $I_1$ is passed by $P_A$ to $S_A$, which results in $O_1$.
        This intermediate output $O_1$ is passed by $P_A$ to $S_B$ as an input, in addition to $I_2$, which results in the outputs $O_2$ and $O_3$.
        The latter are returned by $P_A$ as the final outputs together with the $O_1$ intermediate result.
    }
    \label{fig:schematic-scoped-provenance}
\end{figure*}

It is commonly accepted that science is facing a reproducibility crisis in that many studies can often not be reproduced~\cite{Stoddart:2016}.
In recent years, guidelines have been developed to address this problem, such as the FAIR principles~\cite{Wilkinson:2016} that aim to make data, among other things, more reusable.
For workflows to become FAIR as well, it is critical that they store the \emph{provenance} of the data that they produce at each execution~\cite{Goble:2020}.
Concretely, this means that a workflow should store not only its own inputs and outputs but also those of all the subprocesses that it invokes.
Recording the provenance of data that is produced at each step of a workflow is crucial to enable the reproducibility and intelligibility of the final result.
However, the full provenance is not always required.
Therefore, for complex workflows that produce large provenance graphs, it becomes important to be able to investigate the provenance within different granularity levels, i.e. different scopes.
We refer to the possibility of inspecting the provenance at different levels as scoped provenance, which we illustrate in \cfref{fig:schematic-scoped-provenance}.

The next section explains in detail how we put the concepts of optional transparency and scoped provenance into practice.

\subsubsection{Common workflow design}
To ensure that the common workflows satisfy the requirements of optional transparency and scoped provenance, we have chosen to implement them using AiiDA~\cite{Huber:2020}, a scalable computational infrastructure for automated reproducible workflows and data provenance.
The workflows are implemented as AiiDA \emph{work chains}~\cite{Uhrin:2021}, whose data provenance and that of all their subprocesses is automatically stored by AiiDA in a relational database.
AiiDA provides an \gls{api} to query the provenance graph at various levels of granularity, satisfying the scoped provenance requirement.
The optional transparency criterion is made possible by the design of AiiDA's workflow language specification~\cite{Uhrin:2021}.
All processes in AiiDA are implemented in Python and, most importantly, the process specification (see \clref{code:expose_inputs}) is defined programmatically, allowing inspection of inputs and outputs before executing the workflow.
In addition, it allows workflows to easily reuse subworkflows as modular blocks, without making their interface inaccessible, by \emph{exposing} the inputs and outputs~\cite{Uhrin:2021,Gresch:2018}.

\begin{lstlisting}[
    label=code:expose_inputs,
    numbers=left,
    caption={
        The definition of a process \texttt{ProcessA} implemented as a subclass of an AiiDA \texttt{WorkChain}.
        The process runs two subprocesses (\texttt{SubProcessA} and \texttt{SubProcessB}).
        The process declares an input \texttt{I\_1} in its specification; in addition, the inputs of subprocesses are not redefined, but \texttt{ProcessA} simply exposes them in its own specification.
        The inputs of the subprocesses are exposed in separate namespaces so that inputs with same name do not shadow each other and remain all accessible (see \clref{code:submit_process} for an example of how these are passed).
    }
]
class ProcessA(WorkChain):

    @classmethod
    def define(cls, spec):
        spec.input('I_1')
        spec.expose_inputs(SubProcessA, namespace='S_A')
        spec.expose_inputs(SubProcessB, namespace='S_B')

class SubProcessA(WorkChain):

    @classmethod
    def define(cls, spec):
        spec.input('I_1')
        spec.input('I_2')

class SubProcessB(WorkChain):

    @classmethod
    def define(cls, spec):
        spec.input('I_1')
        spec.input('I_2')
\end{lstlisting}

Launching a process in AiiDA is performed by passing the process class as an argument to the \texttt{submit} function and passing the inputs as keyword arguments as shown in \clref{code:submit_subprocess}.

\begin{lstlisting}[
    label=code:submit_subprocess,
    numbers=left,
    caption={
        Example of how \texttt{SubProcessA} is launched.
        The \texttt{**} marker is Python syntactic sugar to unwrap the \texttt{inputs} dictionary into keyword arguments to the \texttt{submit} function.
        Note that the values of the inputs are simple integers just for clarity of the example.
    }
]
from aiida.engine import submit
inputs = {
    'I_1': 1,
    'I_2': 2,
}
submit(SubProcessA, **inputs)
\end{lstlisting}

\clref{code:submit_process} shows an example of how the top-level \texttt{ProcessA} can be launched, defining its own inputs as well as those of its subprocesses.

\begin{lstlisting}[
    label=code:submit_process,
    numbers=left,
    caption={
        Example of how \texttt{ProcessA} is launched.
        The inputs of the subprocesses can be passed in dictionaries that are nested in the main inputs dictionary, where the keys correspond to the namespace in which the inputs are exposed in the process specification (see \clref{code:expose_inputs}).
    }
]
from aiida.engine import submit
inputs = {
    'I_1': 1,
    'S_A': {
        'I_1': 1,
        'I_2': 2,
    },
    'S_B': {
        'I_1': 1,
        'I_2': 2,
    }
}
submit(ProcessA, **inputs)
\end{lstlisting}

The concept of exposing inputs of subprocesses ensures that the inputs of any subprocess can be controlled from the top-level workflow, regardless of the level of nesting.
This directly satisfies the requirement of providing an opaque interface for expert users that need maximal control.
However, the interface quickly risks becoming complex, as multiply layered workflows will require deeply nested input dictionaries.
The workflow needs to optionally provide a transparent version of the interface to enable also non-expert users to easily use the workflow.

To solve this issue for the common workflows, we implement an input generator for each workflow.
Input generators are not a native AiiDA concept but are a design pattern that emerged from the needs of defining and developing common workflows.
Each common workflow defines a class method \texttt{get\_input\_generator} that returns an instance of an object that acts as the input generator.
The input generator in turn defines the class method \texttt{get\_builder}, which implements the common input interface and returns an instance of a `builder'.
A builder is simply a container that wraps the generated inputs with additional information (such as the workflow class it pertains to), together with additional convenience functionality such as automatic input validation.

Since processes in AiiDA are implemented and executed directly in Python, their functionality can be easily extended.
In addition, by being implemented in the same Python class as the workflow for which the inputs are generated, it is straightforward to keep the two aligned during workflow development.
The \texttt{get\_builder} method of the input generator takes a minimal amount of required arguments and returns a complete set of inputs for the corresponding workflow.
\clref{code:submit_input_generator} shows how the input generator simplifies the usage of \texttt{ProcessA} for users, as now they only need to define a single input.

\begin{lstlisting}[
    label=code:submit_input_generator,
    numbers=left,
    caption={
        Example of how the launching of \texttt{ProcessA} is simplified by generating the inputs through the input generator.
        The \texttt{get\_input\_generator} class method returns an instance whose \texttt{get\_builder} method can be called, to obtain a fully defined builder based on just a single input \texttt{I\_1}.
        The builder, containing all the required inputs, can then be passed directly to the \texttt{submit} function.
        Since the builder also contains the process class for which it is defined, the process class itself no longer has to be explicitly passed to the \texttt{submit} function.
        Here, we are assuming that the inputs of the subprocess can be automatically determined by some algorithm implemented in the input generator; the number of minimally required inputs in this example is just one for simplicity.
    }
]
from aiida.engine import submit
builder = ProcessA.get_input_generator().get_builder(I_1=1)
submit(builder)
\end{lstlisting}

An alternative approach to the problem could have been to make all subprocess inputs optional and let the workflow generate them at runtime (see \cfref{fig:schematic-process-interface}a).
Although this is a valid approach, in our experience it turns out to be much less flexible in particular for experienced users, as internal parameters cannot be set from the outside.
Indeed, the input generator not only makes complex workflows accessible to non-experts, but it also gives maximal flexibility to advanced users.
By generating inputs before execution, they can still be modified according to the user's needs before they are passed to the workflow for execution, giving direct access to all parameters and achieving the goals of optional transparency.

\subsection{The common relax workflow}
\label{sec:common_relax_wf}
As a proof of concept of the principles explained in \csref{sec:reusability}, we present a common interface to a workflow that performs a geometry optimization of both molecular and extended systems, which is implemented for eleven quantum engines.
Structural relaxation towards the most energetically favorable configuration is a common task in materials science, and all selected quantum engines can perform it.
Nevertheless, the quantum engines use a wide variety of algorithms to optimize forces on the atoms and stress on the cell.
This, therefore, presents an ideal yet challenging test scenario to develop a workflow with a common interface.

\subsubsection{Design strategy}\label{sec:design_strat}
\begin{figure}[ht!]
    \centering
    \includegraphics[width=0.9\linewidth]{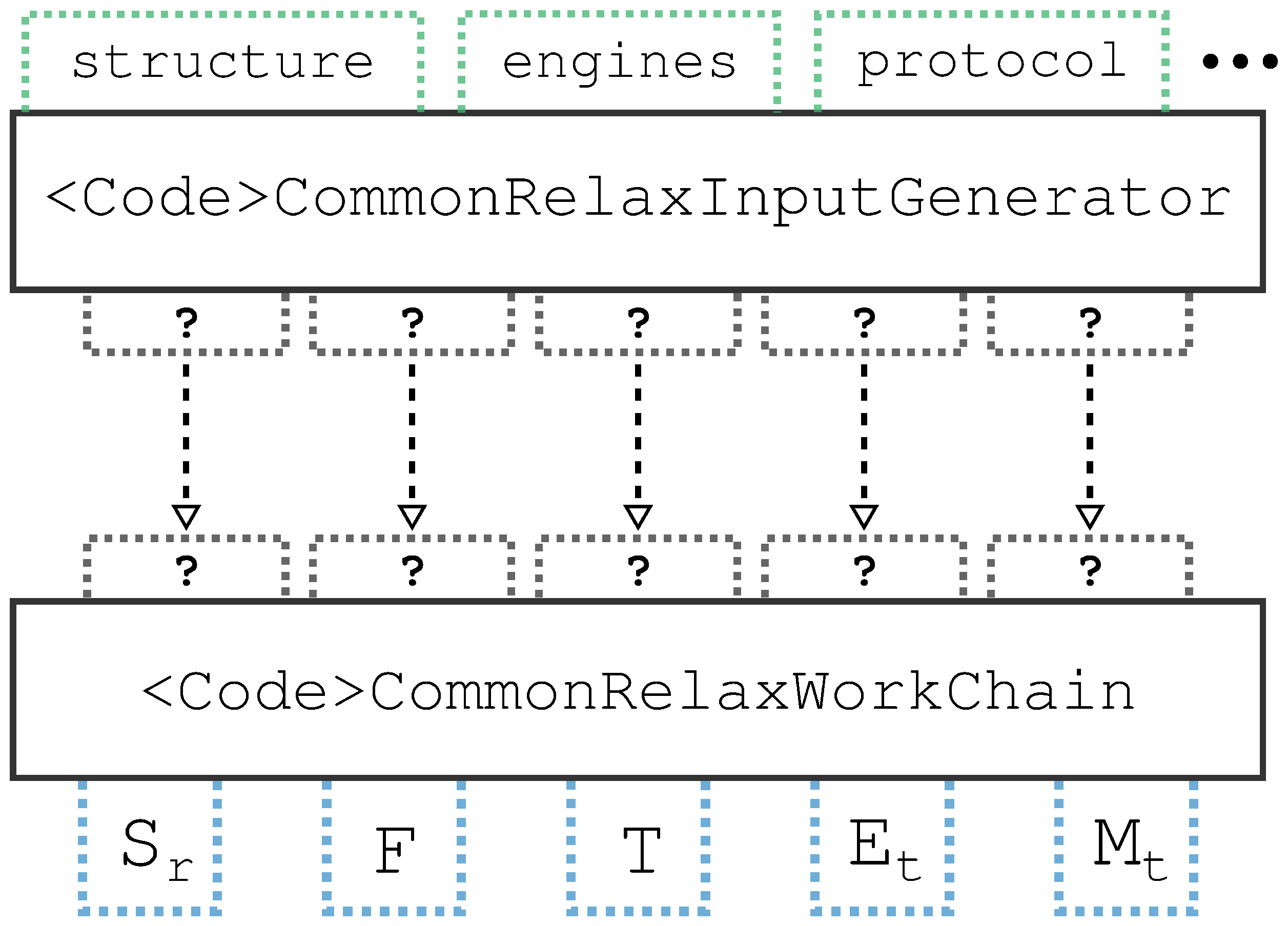}
    \caption{
        {\bfseries Schematic diagram of the common relax workflow interface.}
        Any implementation consists of two parts: the \texttt{<Code>CommonRelaxWorkChain} and a \texttt{<Code>CommonRelaxInputGenerator}.
        The \texttt{<Code>CommonRelaxWorkChain} is an AiiDA \texttt{WorkChain} that implements the logic necessary to perform the structure optimization and has an input interface that is code-specific.
        However, the outputs that it returns respect the schema of the common interface, where $S_r$ is the relaxed structure, $F$ are the forces on each atom, $T$ is the stress on the cell, $E_t$ is the total energy and $M_t$ is the total magnetization of the system.
        Each \texttt{<Code>CommonRelaxWorkChain} provides its own \texttt{<Code>CommonRelaxInputGenerator} which, unlike the workflow, implements the common interface for the inputs (note that not all common inputs are shown for clarity).
        Here \texttt{structure} is the structure that is to be optimized, the \texttt{protocol} is a string that defines how the inputs are determined, and the \texttt{engine} is a dictionary that specifies what code(s) to use.
        The \texttt{<Code>CommonRelaxInputGenerator} translates the common inputs into the code-specific inputs that the corresponding \texttt{<Code>CommonRelaxWorkChain} expects (indicated with \texttt{?}).
        Since the creation of the code-specific inputs and the launching of the workflow are two separate action, the generated inputs can be adapted at will.
    }
    \label{fig:schematic-common-relax-workflow}
\end{figure}

The design of the interface is guided by the idea to employ \emph{optional transparency} to create a workflow interface that is suitable both for expert and non-expert users.
This interface must be simple and general (code-agnostic) but at the same time retain full flexibility in changing code-specific parameters.
Our adopted solution consists in the creation of code-specific workflows (implemented as AiiDA work chains named \texttt{<Code>CommonRelaxWorkChain}, where \texttt{<Code>} indicates the name of the underlying quantum engine), whose interface design is not restricted.
A common interface is achieved by ensuring that each work chain provides also an input generator whose interface is identical for every \texttt{<Code>}, as shown in \cfref{fig:schematic-common-relax-workflow}.

\clref{code:relax_submission} shows an actual code example of how the input generator can be obtained from a work chain implementation.
The \texttt{get\_builder} method of the input generator  will transform the inputs, that respect the common interface, into the inputs that are expected by the corresponding code-specific work chain implementation.
The inputs are returned in the form of a ``builder'' which can be directly submitted to AiiDA to start running the workflow.

\begin{lstlisting}[
    label=code:relax_submission,
    numbers=left,
    caption={
        Submission of the relax common workflow employing the quantum engine \texttt{<Code>}.
        The arguments accepted by \texttt{get\_builder} are identical for every \texttt{<Code>}, establishing a common interface.
    }
]
from aiida.engine import submit
generator = <Code>CommonRelaxWorkChain.get_input_generator()
builder = generator.get_builder(structure=...,protocol=..., ...)
submit(builder)
\end{lstlisting}

We note that not only the names but also the (Python) types of the inputs and outputs are standardized to ensure that the interface is truly generic.
These are described in the next sections.

\subsubsection{Common inputs}\label{sec:common-inputs}
The second step of the design is the identification of the minimal set of arguments for the input generator, reflecting the inputs of the most generic relaxation process.
We identified three fundamental inputs that we therefore implemented as mandatory arguments of the \texttt{get\_builder} method.
\begin{itemize}[leftmargin=*]
    \item \texttt{structure}.
    The structure to relax.
    (type: an AiiDA \texttt{StructureData} instance, the common data format to specify crystal structures and molecules in AiiDA~\cite{Pizzi:2016}).
    \item \texttt{protocol}.
    In the context of this work, this means a single string summarizing the computational accuracy of the underlying \gls{dft} calculation and relaxation algorithm.
    Three protocol names are defined and implemented for each code: \texttt{`fast'}, \texttt{`moderate'} and \texttt{`precise'}.
    The details of how each implementation translates a protocol string into a choice of parameters is code dependent, or more specifically it depends on the implementation choices of the corresponding AiiDA plugin.
    For this work we have tried to follow these definitions: a possibly unconverged (but still meaningful) run that executes rapidly for testing (\texttt{`fast'}); a safe choice for prototyping and preliminary studies (\texttt{`moderate'}); and a set of converged parameters that might result in an expensive simulation but provides converged results (\texttt{`precise'}).
    The reason for not mandating the details of the protocols in the common-workflow specifications is due to the variety of basis sets, input potentials and algorithms, requiring the specification of diverse and heterogeneous parameters in different codes.
    For the eleven implementations presented in this work, we report in the supplementary material the detailed parameter choices and the translation done for each respective code.
    We note here that the choice of the exchange-correlation functional could, in the future, become an additional optional input.
    In this work, we decided to use the \gls{pbe}\cite{Perdew:1996} functional as the default choice.
    (type: a Python string).
    \item \texttt{relax\_type}.
    The type of relaxation to perform, ranging from the relaxation of only atomic coordinates to the full cell relaxation for extended systems.
    The complete list of supported options is: \texttt{`none'}, \texttt{`positions'}, \texttt{`volume'}, \texttt{`shape'}, \texttt{`cell'}, \texttt{`positions\_cell'}, \texttt{`positions\_volume'}, \texttt{`positions\_shape'}.
    Each name indicates the physical quantities allowed to relax.
    For instance, \texttt{`positions\_shape'} corresponds to a relaxation where both the shape of the cell and the atomic coordinates are relaxed, but not the volume; in other words, this option indicates a geometric optimization at constant volume.
    On the other hand, the \texttt{`shape'} option designates a situation when the shape of the cell is relaxed and the atomic coordinates are re-scaled following the variation of the cell, not following a force minimization process.
    The term ``cell'' is short-hand for the combination of \texttt{`shape'} and \texttt{`volume'}.
    The option \texttt{`none'} indicates the possibility to calculate the total energy of the system without optimizing the structure.
    Not all the described options are supported by each code involved in this work; only the options \texttt{`none'} and  \texttt{`positions'} are shared by all the eleven codes.
    The supported options might be extended in the future.
    (type: a Python string).
\end{itemize}
In addition to these mandatory arguments, the computational resources must be passed to the work chain in order to make the interface transferable between different computational environments.
For this task, a specific argument of \texttt{get\_builder} has been designed called \texttt{engines}.
\begin{itemize}[leftmargin=*]
    \item \texttt{engines}. It specifies the codes and the corresponding computational resources for each step of the relaxation process.
    Typically one single executable is sufficient to perform the relaxation.
    However, there are cases in which two or more codes in the same simulation package are required to achieve the final goal, as for example in the case of \fleur{}.
    (type: a Python dictionary).
\end{itemize}
Other inputs have been recognized as common optional features that also a non-expert user might want to have control over:
\begin{itemize}[leftmargin=*]
    \item \texttt{threshold\_forces}.
    A real positive number indicating the target threshold for the forces in eV/{\AA}.
    If not specified, the protocol specification will select an appropriate value.
    (type: Python float).
    \item \texttt{threshold\_stress}.
    A real positive number indicating the target threshold for the stress in eV/{\AA}$^3$.
    If not specified, the protocol specification will select an appropriate value.
    (type: Python float).
    \item \texttt{electronic\_type}.
    An optional string to signal whether to perform the simulation for a metallic or an insulating system.
    It accepts only the \texttt{`insulator'} and \texttt{`metal'} values.
    This input is relevant only for calculations on extended systems.
    In case no such option is specified, the calculation is assumed to be metallic which is the safest assumption.
    (type: Python string).
    \item \texttt{spin\_type}.
    An optional string to specify the spin degree of freedom for the calculation.
    It accepts the values \texttt{`none'} or \texttt{`collinear'}.
    These will be extended in the future to include, for instance, non-collinear magnetism and spin-orbit coupling.
    The default is to run the calculation without spin polarization.
    (type: Python string).
    \item \texttt{magnetization\_per\_site}.
    An input devoted to the initial magnetization specifications.
    It accepts a list where each entry refers to an atomic site in the structure.
    The quantity is passed as the spin polarization in units of electrons, meaning the difference between spin up and spin down electrons for the site.
    This also corresponds to the magnetization of the site in Bohr magnetons ($\mu_B$).
    The default for this input is the Python value \texttt{None} and, in case of calculations with spin, the \texttt{None} value signals that the implementation should automatically decide an appropriate default initial magnetization.
    The implementation of such choice is code-dependent and described in the supplementary material of this manuscript.
    (type: \texttt{None} or a Python list of floats).
    \item \texttt{reference\_workchain}.
    A previously performed \texttt{<Code>CommonRelaxWorkChain}.
    When this input is present, the interface returns a set of inputs which ensure that results of the new \texttt{<Code>CommonRelaxWorkChain} can be directly compared to the \texttt{reference\_workchain}.
    This is necessary to create, for instance, meaningful equations of state.
    Its use will be clarified in the \csref{sec:eos}.
    (type: a previously completed \texttt{<Code>CommonRelaxWorkChain}).
\end{itemize}

The arguments of the input generator described above fully satisfy the needs for the creation of a ``ready-to-submit'' \texttt{<Code>CommonRelaxWorkChain}, constructing all its necessary \texttt{inputs} (see \clref{code:relax_submission}).
These inputs are code-specific and, as discussed earlier, can be modified before submission by an expert user who is familiar with the internals of the \texttt{<Code>CommonRelaxWorkChain}.

\subsubsection{Inspection of valid inputs}\label{sec:inspection-and-guis}
The arguments of the \texttt{get\_builder} method represent high-level parameters that describe how the geometry optimization should be performed or how the system is to be treated.
Each argument has a fixed number of accepted values, but not every code implementation may necessarily support all of them, as some values might correspond to features not supported by the code.
In order to be able to inspect which options are supported by a workflow implementation, the input generator offers a number of methods.
An example is shown in \clref{code:generator_inspectors}.
\begin{lstlisting}[
    label=code:generator_inspectors,
    numbers=left,
    caption={
        Call to the inspection method that returns information on the available relaxation types for the \texttt{<Code>} implementation of the common relax workflow.
    }
]
input_gen = <Code>CommonRelaxWorkChain.get_input_generator()
input_gen.get_relax_types()
\end{lstlisting}
The \texttt{get\_relax\_types} method returns the supported values for \texttt{relax\_type} for the corresponding workflow implementation.
Inspection methods are implemented for all codes and all the arguments of \texttt{get\_builder} except the threshold values, the \texttt{structure}, the \texttt{reference\_workchain}, and the \texttt{magnetization\_per\_site}.
Associated to the \texttt{engines} argument, there are the methods \texttt{get\_engine\_types} and \texttt{get\_engine\_type\_schema}, which return the steps required by the relaxation and information on the code type necessary for each step of the relaxation, respectively.

The described inspection methods allow to introspect, in a fully machine-readable and automatic way, what the valid options for a particular common workflow implementation are.
This is particularly relevant to facilitate future development of a \gls{gui} for the submission of the common relax workflow.
The \gls{gui} will be able to create the necessary input fields, with a list of accepted values, by programmatically introspecting the input generator interface.

\subsubsection{Common outputs}\label{sec:common-outputs}
To allow direct comparison and cross-verification of the results, the outputs of \texttt{<Code>CommonRelaxWorkChain} are standardized for all implementations and are defined as follows:

\begin{itemize}[leftmargin=*]
    \item \texttt{forces}.
    The final forces on all atoms in eV/{\AA}.
    (type: an AiiDA \texttt{ArrayData} of shape $N\times 3$, where $N$ is the number of atoms in the structure).

    \item \texttt{relaxed\_structure}.
    The structure obtained after the relaxation.
    It is not returned if the \texttt{relax\_type} is \texttt{`none'}.
    (type: AiiDA's \texttt{StructureData}).

    \item \texttt{total\_energy}.
    The total energy in eV associated to the relaxed structure (or initial structure in case no relaxation is performed).
    The total energy is not necessarily defined in a code-independent way (e.g., it does not have a common zero).
    We require, however, that the partial derivative of the returned energy with respect to the change of the coordinate $i$ of atom $j$ is always the $i-$th coordinate of the force on the atom $j$.
    We also stress that in general, even for calculations performed with the same code, there is no guarantee to have comparable energies in different runs if the inputs are generated with the input generator (because, for instance, the selected k-points depend on the input structure volume).
    However, in combination with the input argument \texttt{reference\_workchain} mentioned earlier and discussed in \csref{sec:eos}, energies from different relaxation runs become comparable, and their energy difference is well defined.
    (type: AiiDA \texttt{Float}).

    \item \texttt{stress}.
    The final stress tensor in eV/{\AA}$^3$.
    Returned only when a variable-cell relaxation is performed.
    (type: AiiDA \texttt{Float}).

    \item \texttt{total\_magnetization}.
    The total magnetization in $\mu_B$ (Bohr-magneton) units.
    Returned only for magnetic calculations.
    (type: AiiDA \texttt{Float}).
\end{itemize}

\subsubsection{A simple test case: ammonia}
As a first test case of the various implementations of the common relax workflow, we present the optimization of a simple molecular structure: ammonia.
The thermodynamically stable polymorph of ammonia has a trigonal pyramidal shape, which makes the structure polar.
However, ammonia also exists in a metastable planar form \cite{Ghosh:2000}.
The optimized structure and its associated total energy have been calculated with the common relax workflow implementation for all eleven quantum engines discussed in this paper for both phases of ammonia, using the \texttt{`precise'} protocol for the input generation.

\begin{figure}[ht]
    \centering
    \resizebox{0.99\linewidth}{!}{\input{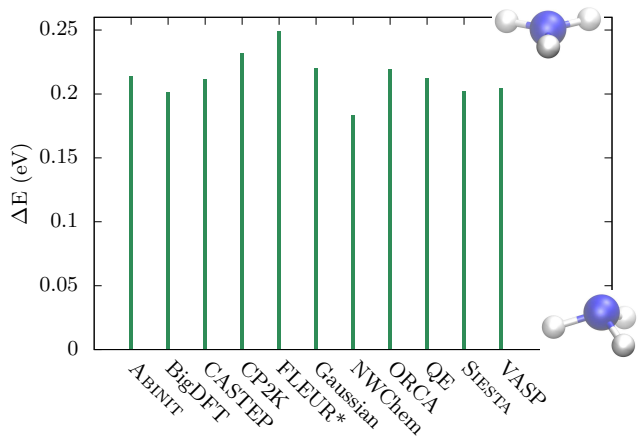}}
    \caption{
        {\bfseries Energy difference between the planar and pyramidal phases of ammonia.}
        Energy difference ($\Delta$E) between the planar and pyramidal phases of ammonia, calculated with the eleven quantum engines reported in the horizontal axis (QE stands for \qe{}).
        A relaxation of the structure has been performed independently by every code before computing the energies (except for \fleur{}$^*$, because for \fleur{} the relaxation failed due to overlapping muffin-tin spheres, a method-specific issue requiring error handling.
        The energy difference for \fleur{} was calculated through the common workflow without relaxation, using the relaxed output structures of \qe{} instead).
    }
    \label{fig:ammonia}
\end{figure}

The analysis of the energy difference between the planar and pyramidal configurations of ammonia is presented in \cfref{fig:ammonia}.
As mentioned in the introduction, comparing results among codes is not the focus of this paper.
However, it is worth mentioning that the small discrepancies between codes in \cfref{fig:ammonia} are not surprising, considering that the treatment of polar molecules with codes designed for extended systems is not a trivial task.
In particular, some codes always need to use periodic boundary conditions, introducing non-physical interactions among replicas in the calculation.
Even for large enough simulation cells, the long-range electrostatic potential due to periodic images of the polar molecule affects the energy of the system.
Strategies such as the use of improved Poisson solvers \cite{Cerioni:2012,Castro:2003} and more sophisticated dipole corrections \cite{Bengtsson:1999,Makov:1995} can be introduced in order to circumvent this problem.
Since in this paper we are focusing only on showing the concept and feasibility of common workflows, no dipole correction is considered, and the simulation box is set to a (15 {\AA})$^3$ cube, without performing a proper convergence study on the cell size.
However, extensions of this work can add optional flags to the input generator of the workflow to activate appropriate dipole corrections if needed and implemented by the underlying quantum engine.
The data presented here also illustrates the potential of the common interface for the cross-verification of results, especially considering the variety of basis sets and algorithms of the eleven quantum engines.
The present work offers the possibility to compare results from quantum-chemistry-oriented and electronic-structure codes (both pseudopotentials- and all-electrons-based) with minimum effort.

\subsubsection{Provenance}
\begin{figure}
    \centering
    \includegraphics[width=0.8\columnwidth]{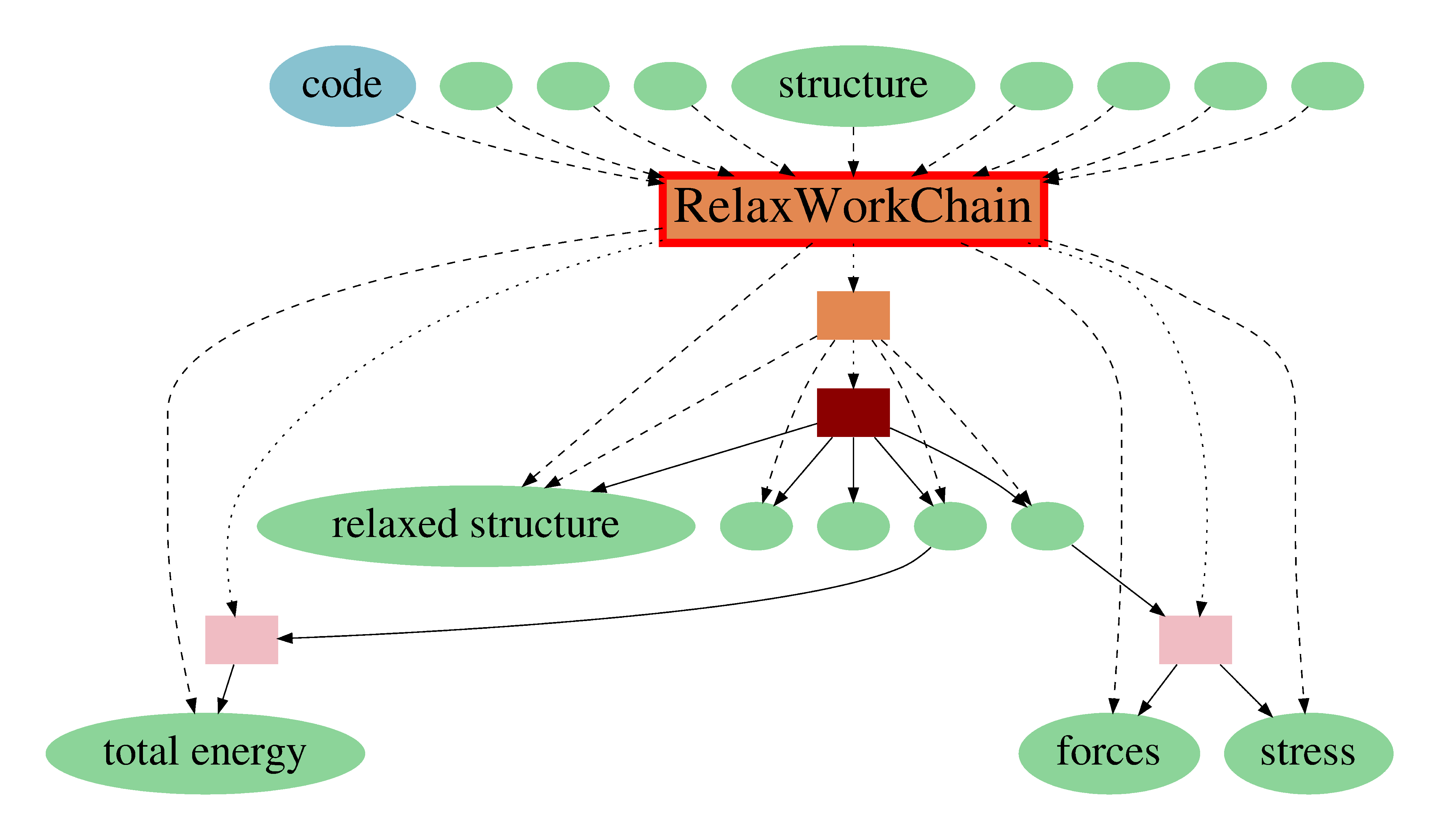}
    \includegraphics[width=0.99\columnwidth]{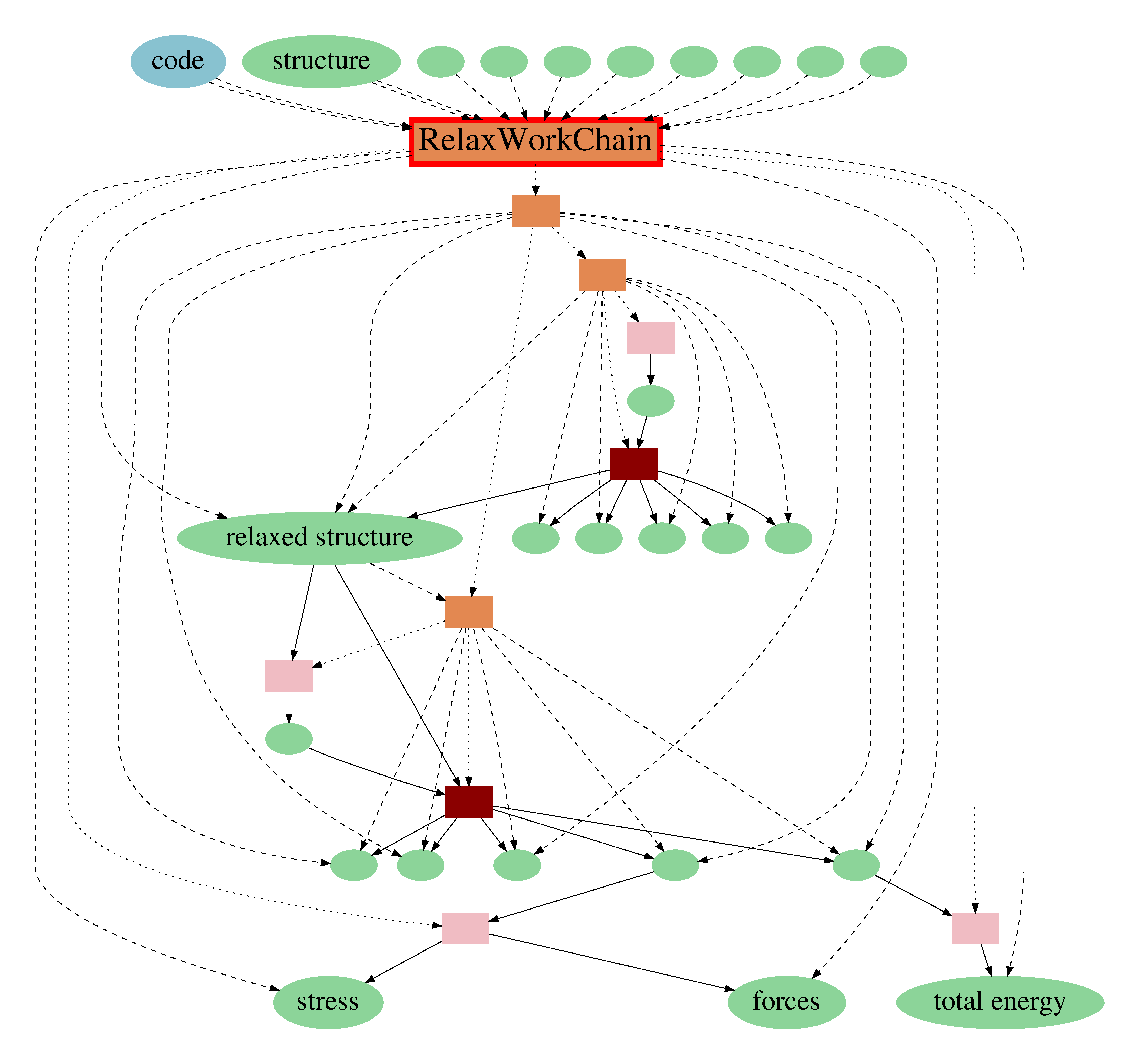}
    \caption{
        {\bfseries Schematic provenance graph for a relaxation workflows.}
        Schematic provenance graph for a relaxation workflow powered by two different quantum engines (top: \siesta{}; bottom: \qe{}).
        The node of the \texttt{<Code>CommonRelaxWorkChain} is highlighted with the label ``RelaxWorkChain'' and with a red edge.
        All the AiiDA work chains called during the relaxation are represented by orange rectangles.
        The dark red rectangles are calculations, meaning calls to an external executable that performs a calculation, for instance a call to the \texttt{pw.x} of \qe{}.
        All the ellipses represents data nodes, meaning nodes in the database that contain data, like, for instance, the initial structure, the total energy and so on. In blue is the data node representing the code utilized for the calculations. In pink are represented calls to Python functions that modify some data in order to create others.
    }
    \label{fig:prov-relax}
\end{figure}

Since all workflows are implemented using AiiDA, the full provenance is automatically stored when the workflow is executed, as discussed in \csref{sec:reusability}.
\cfref{fig:prov-relax} shows a schematic provenance graph for a relaxation workflow powered by two different quantum engines.
Note that only a subselection of the total number of inputs and outputs are shown for clarity, but all subprocesses are displayed and the connections between the nodes give an idea of the internal complexity of the workflows.
Notably, the figure shows how the different workflow implementations can follow considerably different logical paths while ultimating returning the same quantities according to the same common interface.

The action of taking the arguments of the common interface and transforming them in code-specific inputs (operated by the input generators) is not tracked.
This is not crucial since only the code specific inputs fully determine the calculation results.
Also, we do not consider protocols as immutable objects, but rather as flexible input suggestions that an expert user might want to change.

\subsection{Code-agnostic workflows}
Optimizing the geometry of a solid-state structure or molecule is a common core building block of materials-science workflows.
Creating a common workflow interface for this particular task allows higher level workflows that reuse it to become code agnostic.
This makes it possible to run the workflow with any quantum engine that implements the common relax workflow interface, without having to explicitly specify any input that is specific to the quantum engine.
We discuss two examples of such workflows in the following sections.

\subsubsection{Equation of State}\label{sec:eos}
\begin{figure}
    \centering
    \includegraphics[width=0.95\columnwidth]{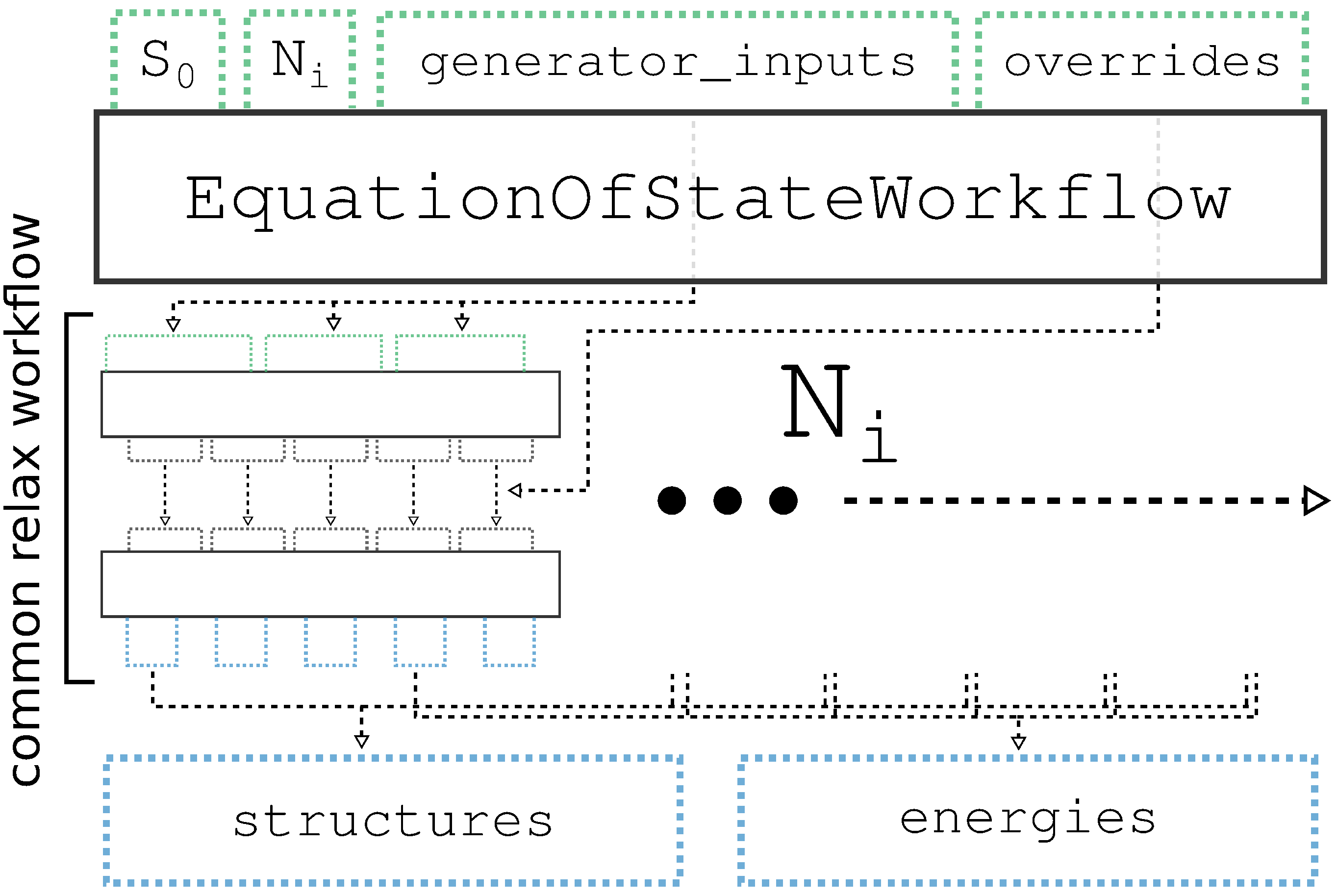}
    \caption{
        {\bfseries Schematic diagram of the code-agnostic \gls{eos} workflow.}
        Schematic diagram of the implementation of the code-agnostic \texttt{EquationOfStateWorkflow}.
        The \texttt{EquationOfStateWorkflow} takes a number of arguments: \texttt{$S_0$} is the structure of the system at equilibrium volume and \texttt{$N_i$} are the number of volumes for which to compute the total energy.
        The \texttt{generator\_inputs} will be passed directly to the inputs generator of the chosen common relax workflow implementation, which is called \texttt{$N_i$} times, once for each system volume.
        Note that the inset marked as ``common relax workflow'' corresponds directly to the schematic of the common relax workflow in \cfref{fig:schematic-common-relax-workflow}.
        This highlights that the \texttt{EquationOfStateWorkflow} directly reuses the common relax workflow as its main building block.
        Which implementation of the common relax workflow is to be used is communicated to the \texttt{EquationOfStateWorkflow} by a single input, which is not shown for clarity.
        The \texttt{overrides} port allows an expert user to override certain inputs that are automatically determined by the generator, thus making the \texttt{EquationOfStateWorkflow} optionally transparent.
    }
    \label{fig:schematic-code-agnostic-workflow}
\end{figure}

An example of a workflow that uses structure optimization as a building block is the \gls{eos} workflow.
The equation of state of a solid-state system is obtained by computing the total energy of the system at various volumes.
We present here the implementation of an \gls{eos} workflow that uses the common relax workflow to perform the optimization of the system at each volume and compute its total energy.
This workflow serves as an example to explain how the unified interface of the common relax workflow can be used to create code-agnostic workflows.
It has been named \texttt{EquationOfStateWorkflow} and its schematic representation is shown in \cfref{fig:schematic-code-agnostic-workflow}.

The \texttt{EquationOfStateWorkflow} takes a structure as input ($S_0$ in \cfref{fig:schematic-code-agnostic-workflow}) and scales the volume a number of times ($N_i$), with the scaled structures centered around the volume of the input structure.
The workflow calls the common relax workflow for each scaled structure to compute its total energy.
The common relax workflow interface is entirely accessible at the level of the inputs of the \texttt{EquationOfStateWorkflow}.
This means that one can specify arguments accepted by the input generator (which are code-agnostic and are labeled \texttt{generator\_inputs} in \cfref{fig:schematic-code-agnostic-workflow}), but also, optionally, some code-specific \texttt{overrides} for the inputs produced by the generator.
Therefore, on the one hand, by virtue of the common interface being code-agnostic, the \texttt{EquationOfStateWorkflow} is also independent of the quantum engine that is used for the underlying calculations.
On the other hand, the possibility to specify explicit \texttt{overrides} should fulfill the needs of expert users and fully satisfy the optional transparency requirement for reusable workflows.

The total energies and optimized structure, as produced by the common relax workflow runs, are collected and returned by the \texttt{EquationOfStateWorkflow} as its outputs.
Like the \texttt{<Code>CommonRelaxWorkChain} itself, the code-agnostic \texttt{EquationOfStateWorkflow} is implemented as an AiiDA work chain.
This provides fully automated provenance tracking of all tasks performed inside the workflow, ensuring full reproducibility of the computed results.

It should be noted that the actual logic of the \texttt{EquationOfStateWorkflow} is slightly more complicated than depicted in \cfref{fig:schematic-code-agnostic-workflow}.
The common relax workflows are not all launched in parallel, but a single workflow is first performed for one of the scaled volumes.
This first workflow is subsequently used as an additional input for the \texttt{reference\_workchain} argument to the input generator for the common relax workflows for the remaining volumes.
The input generator can use this reference to the first workflow to ensure that, if needed, parameters are kept constant between images in order for the energy differences to be meaningful.
An example is the number of k-points used to sample the Brillouin zone (that is typically chosen by the input generators so as to get as close as possible to a target density, and thus is volume-dependent if a \texttt{reference\_workchain} is not specified).

The \texttt{EquationOfStateWorkflow} has been used to compute the \gls{eos} for a number of solid-state systems with varying electronic and magnetic properties: silicon (Si), aluminium (Al), germanium telluride (GeTe) and \gls{bcc} iron (Fe) both in a ferromagnetic and anti-ferromagnetic configuration.
The results are shown in \cfref{fig:Eos1} and \cfref{fig:Fe}.

\begin{figure*}
    \centering
    \resizebox{0.99\linewidth}{!}{\input{images/plot_eos.tex}}
    \caption{
        {\bfseries \gls{eos} for Si, Al and GeTe.}
        Results obtained with the code-agnostic \texttt{EquationOfStateWorkflow}.
        For each code, the energy is shifted to set the minimum energy to zero.
        The EOS has been computed with all codes discussed in this work, except \orca{} and \gaussian{}, which are mainly specialized for non-periodic systems.
        In addition, for GeTe, results are missing for \bigdft{}, \cptwok{}, \fleur{} and \nwchem{} (see Table II in the Supplementary Information for more details).
        The label QE stands for \qe{}.
    }
    \label{fig:Eos1}
\end{figure*}

\begin{figure*}
    \centering
    \resizebox{0.99\linewidth}{!}{\input{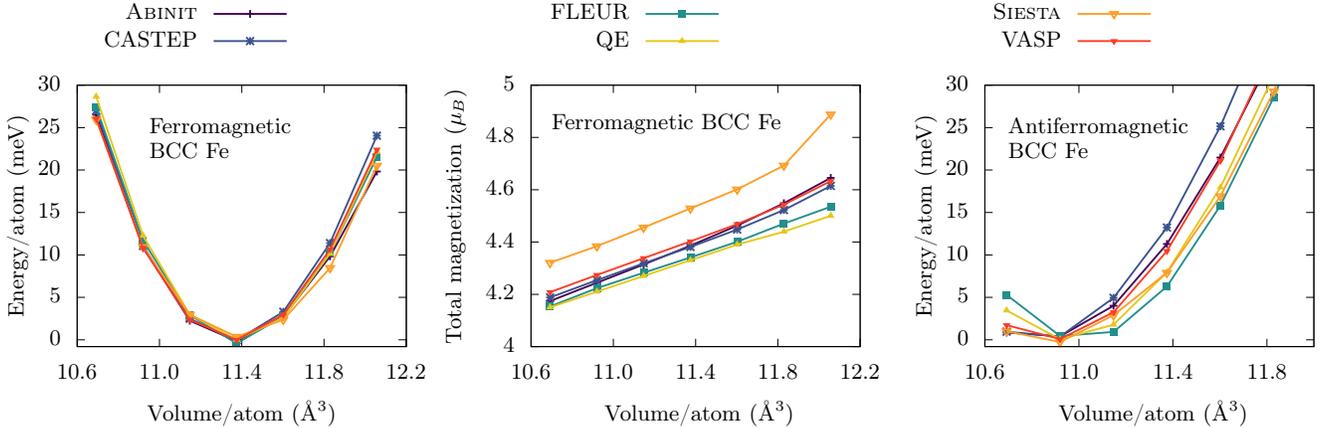}}
    \caption{
        {\bfseries\gls{eos} and total magnetization for \gls{bcc} Fe.}
        Results obtained with the code-agnostic \texttt{EquationOfStateWorkflow}.
        The left panel reports the \gls{eos} obtained with a ferromagnetic initialization of the atomic moments.
        The corresponding total magnetization is reported for each volume in the central panel.
        The right panel reports the \gls{eos} obtained with an anti-ferromagnetic initialization of the atomic moments.
	    The label QE stands for \qe{}.
	    Results are missing for \bigdft{}, \cptwok{}, \gaussian{}, \nwchem{} and \orca{} (see Table II in the Supplementary Information for more details).
    }
    \label{fig:Fe}
\end{figure*}

\cfref{fig:Eos1} reports the \gls{eos} results for the Si, Al and GeTe crystals.
The curves for Si and Al have been obtained with all quantum engines, except \orca{} and \gaussian{}, which are mainly specialized for non-periodic systems and the \gaussian{} AiiDA plugin does not yet support PBC.
At each volume, the atomic positions are optimized while keeping the volume and cell shape fixed.
The GeTe compound crystallizes at normal conditions in a trigonal phase (space group $R3m$) \cite{Goldak:1966}.
For this material, a correct calculation of the \gls{eos} requires the cell shape to be optimized at fixed volume in order to minimize the non-hydrostatic contributions of the stress tensor.
This has been achieved in the common interface setting the \texttt{relax\_type} to \texttt{positions\_shape} (see \csref{sec:common-inputs}), which is only supported by five out of eleven quantum engines.
This is the reason why only five curves are shown in the right panel of \cfref{fig:Eos1}.
All calculations are carried out without spin-polarization and with the \texttt{precise} protocol.

The common interface also allows calculations on magnetic systems.
\cfref{fig:Fe} shows the \gls{eos} of \gls{bcc} Fe, for both a ferromagnetic (left panel) and anti-ferromagnetic (right panel) ordering of atomic spin moments.
At each volume, the atomic positions are optimized while keeping the volume and cell shape fixed.
The central panel in \cfref{fig:Fe} shows the total magnetization of the relaxed structure at each volume in the ferromagnetic case.
The total magnetization in the anti-ferromagnetic case is zero at every volume and therefore not reported in the picture.
The initial structure passed to the workflow is the same for the ferromagnetic and anti-ferromagnetic configurations and it is close to the equilibrium volume of the ferromagnetic case.
This explains why the volume with minimum energy for the anti-ferromagnetic case is not placed in the middle of the analyzed volumes range.
It is noteworthy that the \gls{bcc} structure is the most thermodynamically stable configuration only in the ferromagnetic arrangement.
The results show good overall agreement among codes.
However, the scope of this section is only to demonstrate the variety of systems and physical quantities that can be analyzed with the code-agnostic \gls{eos} workflow.

\subsubsection{Dissociation curve}
In a similar fashion to the \gls{eos} workflow, a code-agnostic workflow for the calculation of the dissociation curve of a diatomic molecule has been implemented.
In this case, no relaxation is performed at all by the common relax workflow (accomplished by setting the \texttt{relax\_type} equal to \texttt{`none'}) and it simply computes the energy of the system at various atomic distances.
The same approach of the \gls{eos} workflow is used regarding the \texttt{reference\_workchain} argument, meaning that the calculation at the first distance is used as a reference for the creation of inputs for the calculation at all the other distances.

\begin{figure}
    \centering
    \resizebox{0.99\columnwidth}{!}{\input{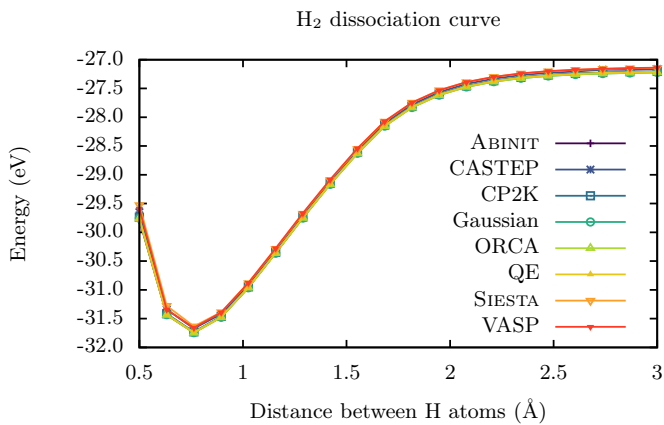}}
    \caption{
        {\bfseries Dissociation curve of the H$_2$ molecule}. Results obtained with a code-agnostic workflow.
        For all codes and all volumes, the magnetization is initialized to -1 $\mu_B$ for one atom and to
        1 $\mu_B$ for the other atom. The label QE stands for \qe{}. The \vasp \- curve has been shifted by -24.89 eV.
        Results are missing for \bigdft{}, \fleur{} and \nwchem{} (see Table II in the Supplementary Information for more details).
    }
    \label{fig:H2}
\end{figure}

Results are presented in \cfref{fig:H2} for the H$_2$ dissociation curve obtained with the code-agnostic workflow.
An initial anti-ferromagnetic configuration has been chosen as a starting point for each energy calculation.
The results show good agreement among codes.
\gls{dft} is not the most appropriate method for the calculation of dissociation curves in diatomic molecules, since these systems expose well known problems of \gls{dft}, like the delocalization error (self-interaction error) and static correlation \cite{Cohen:2008}. The present test case wants to demonstrate the possibility to create code-agnostic workflows that support both electronic-structure codes and quantum-chemistry-oriented codes. In the future, the common relax workflow could be extended to  allow calculations powered by different methods in addition to \gls{dft}, elevating the present work to a useful tool for comparing different levels of theory in the study of crystals and molecules.

\subsection{Conclusions}
We have described how it is possible for domain experts to provide robust and reusable workflows that automatically compute materials properties, in order to exploit the ever-increasing computational power and popularity of \gls{dft}-based quantum engines, with the goal of accelerating materials discovery and characterization.
For the workflows to be reusable, it is critical that they have \emph{optionally transparent} interfaces and that the full provenance of executed workflows is automatically stored.
We have demonstrated a concrete implementation of these two requirements using the workflow management system of the AiiDA informatics infrastructure~\cite{Huber:2020}.
We defined a common interface for a workflow that optimizes the geometry of a solid-state system or molecule, that was subsequently implemented for eleven popular quantum engines, with very diverse basis-set choices and algorithms.
Using this common relax workflow, we have shown how higher-level workflows can reuse it to compute relevant material properties, such as the \gls{eos} and \gls{dc}, while keeping a fully code-agnostic interface.
Our results show how optionally transparent common workflow interfaces directly enable the cross-verification of results produced by different quantum engines.
In addition, they empower a broader audience to use these methods in a robust way, encoding the experts' knowledge into reproducible code and hopefully stimulating new collaborations and more accurate materials science simulations.

\section{Methods}
\subsection{Quantum Mobile implementation}
The common interface and all corresponding code-agnostic workflows described in this paper allow anyone to run calculations to perform the same task with different codes, without knowing the details of each implementation.
This is true assuming that the user can access a working executable of each code.
The executable can be installed on the same machine as AiiDA and the common workflows, or more typically in a remote computer (HPC cluster or supercomputer), since AiiDA allows automatic connection to external machines.
Compiling and installing eleven different quantum engines can be a burden even for experienced users, and even more for non-experts.
As one of the goals of this work is to facilitate the access to quantum codes to a broader audience, we also make available all codes related to this project in Quantum Mobile\cite{Talirz:2020} version 21.05.1, which can be \href{https://quantum-mobile.readthedocs.io/en/latest/releases/versions/21.05.1.html}{downloaded here}.
Quantum Mobile is an open-source virtual machine based on Ubuntu, that comes with a large number of codes, tools and dependencies that are commonly needed to run materials-science atomistic simulations.
In particular, it contains a pre-configured AiiDA installation together with the plugins interfacing AiiDA to all eleven quantum engines described here.
In addition, since version 21.05.1 Quantum Mobile also includes the common-workflow interfaces and implementations discussed here, released as the \href{https://pypi.org/project/aiida-common-workflows/}{\texttt{aiida-common-workflows} package v0.1 on PyPI}.
Crucially, Quantum Mobile also includes the executables for the following open-source quantum engines: \abinit{}, \bigdft{}, \cptwok{}, \fleur{}, \nwchem{}, \qe{} and \siesta{} (as well as a few more).
Although \castep{} and \orca{} provide free academic licenses, it requires users to have their own license which prevents pre-installation in Quantum Mobile.
The remaining three codes discussed here (\castep{}, \gaussian{} and \vasp{}) are commercial, therefore they cannot be redistributed freely without infringing their licenses.
Nevertheless, a complete set of instructions is provided in the supplementary material, to guide users who already have access to these codes (on any computer of their choice) to configure them with AiiDA.
In this way, the common workflows (all instead available open-source in the Quantum Mobile) can be run seamlessly also for commercial codes.
Thanks to this setup, common workflows with these codes can be run with almost no preliminary step required.
Only few codes require minimal adjustment that are described in the subsections \textit{``running in the Quantum Mobile virtual machine''} of the supplementary material.
A detailed list of instructions on how to run the test cases presented in this manuscript in the Quantum Mobile is reported in the Supplementary Material.

\section*{Code availability}
\label{sec:codeavailability}
The source code of the common workflows is released under the MIT open-source license and is made available on GitHub (\href{https://github.com/aiidateam/aiida-common-workflows}{github.com/aiidateam/aiida-common-workflows}).
It is also distributed as an installable package through the Python Package Index (\href{https://pypi.org/project/aiida-common-workflows}{pypi.org/project/aiida-common-workflows}).

\section*{Data availability}
\label{sec:dataavailability}
The data and the scripts used to create all the images in this work are available on the Materials Cloud Archive~\cite{web:Huber:MCA:2021}.
Note that the data includes the entire AiiDA provenance graph of each workflow execution, as well as the curated data that is extracted from that database in order to produce the images.

\section*{Acknowledgments}
This work is supported by the MARVEL National Centre of Competence in Research (NCCR) funded by the Swiss National Science Foundation (grant agreement ID~51NF40-182892) and by the European Union's Horizon 2020 research and innovation programme under grant agreement No.~824143 (European MaX Centre of Excellence ``Materials design at the Exascale'') and grant agreement No.~814487 (INTERSECT project).
The authors thank M. Giantomassi and J.-M. Beuken for their contributions in adding support for \texttt{PseudoDojo} tables to the \href{https://github.com/aiidateam/aiida-pseudo}{\texttt{aiida-pseudo} plugin}. The authors also thank X. Gonze, M. Giantomassi, M. Probert, C. Pickard, P. Hasnip, J. Hutter, M. Iannuzzi, D. Wortmann, S. Blügel, J. Hess, F. Neese and P. Delugas for providing useful feedback on the various quantum engine implementations.
SP acknowledges support from the European Unions Horizon 2020 Research and Innovation Programme, under the Marie Sk\l{}odowska-Curie Grant Agreement SELPH2D No.~839217.
EFL acknowledges the support of the Norwegian Research Council (project number 262339) and computational resources provided by Sigma2.
PZP thanks the Faraday Institution CATMAT project (EP/S003053/1, FIRG016) for financial support.
KE acknowledges the Swiss National Science Foundation (grant number 200020-182015).
GPi and KE acknowledge the swissuniversities ``Materials Cloud'' (project number 201-003).
Work at ICMAB is supported by the Severo Ochoa Centers of Excellence Program (MICINN CEX2019-000917-S), by PGC2018-096955-B-C44 (MCIU/AEI/FEDER, UE), and by GenCat 2017SGR1506.
BZ thanks the Faraday Institution FutureCat project (EP/S003053/1, FIRG017) for financial support.
JB and VT acknowledge support by the Joint Lab Virtual Materials Design (JLVMD) of the Forschungszentrum J\"ulich.

\section*{Author contributions}
GPi and NM conceived the idea of a common interface among quantum engines and supervised the project.
SPH, EB and GPi coordinated the project, implemented the interface design for the common relax workflow and the code-agnostic workflow implementations that use it as a modular block.
SPH, MU and DG conceived the design of reusable modular workflow interfaces through programmatic process specification and implemented it in AiiDA.
CS created the Quantum Mobile virtual machine to include the AiiDA plugins containing the implementation of the common relax workflow interface for all quantum engines, as well as an installation of the quantum engines for those that are open-source and can be freely redistributed.
SP, ZP and GPe developed the \abinit{} implementation of the common relax workflow which relies on the \texttt{aiida-abinit} plugin developed and maintained by SP, AZ, and GPe, and the work was supervised by GMR.
AD developed the \bigdft{} implementation of the common relax workflow which relies on the \texttt{BigDFTCommonRelaxWorkChain} and \texttt{BigDFTBasexorkChain} of the \texttt{aiida-bigdft} plugin package, supervised by LG.
BZ developed the \castep{} implementation of the common relax workflow and its protocols which relies on the \texttt{CastepCommonRelaxWorkChain} of the package \texttt{aiida-castep} developed by the same author.
AVY developed the \cptwok{} implementation of the common relax workflow and its protocols which rely on the \texttt{Cp2kBaseWorkChain} of the \texttt{aiida-cp2k} package, developed by AVY and others, and the work was supervised by BS.
JB developed the \fleur{} implementation of the common relax workflow and its protocols which rely on the \texttt{FleurBaseCommonRelaxWorkChain}, \texttt{FleurCommonRelaxWorkChain}, \texttt{FleurScfWorkChain} and \texttt{FleurBaseWorkChain} of the \texttt{aiida-fleur} package\cite{Broeder:2019}, developed by JB, VT, DW and others, under the supervision of DW.
KE developed the \gaussian{} implementation of the common relax workflow which relies on the \texttt{GaussianBaseWorkChain} of the package \texttt{aiida-gaussian}, developed by KE and PZP.
CJ developed the \nwchem{} implementation of the common relax workflow and its protocols which relies on the \texttt{NwchemBaseWorkChain} of the package \texttt{aiida-nwchem}, developed by CJ and others.
PZP developed the \orca{} implementation of the common relax workflow which relies on the \texttt{OrcaBaseWorkChain} of the \texttt{aiida-orca} package developed by the same author.
SPH developed the \qe{} implementation of the common relax workflow with the support of MB, which relies on the \texttt{PwCommonRelaxWorkChain} and \texttt{PwBaseWorkChain} of the \texttt{aiida-quantumespresso} package, developed by SPH, MB, GPi and others.
EB developed the \siesta{} implementation of the common relax workflow which relies on the \texttt{SiestaBaseWorkChain} of the package \texttt{aiida-siesta}\cite{Garcia:2020}, developed by AG, VD and EB.
EFL developed the \vasp{} implementation of the common relax workflow which relies on the \texttt{CommonRelaxWorkChain}, \texttt{VerifyWorkChain} and \texttt{VaspWorkChain} of the AiiDA plugin \texttt{aiida-vasp}, developed by EFL and others.

\section*{Competing interests}
The authors declare no competing interests.

\clearpage

\onecolumngrid
\begin{center}
  \textbf{\large Supplementary Information: Common workflows for computing material properties using different quantum engines}\\[.2cm]
(Dated: \today)\\[1cm]
\end{center}

\setcounter{equation}{0}
\setcounter{figure}{0}
\setcounter{table}{0}
\setcounter{page}{1}
\renewcommand{\theequation}{S\arabic{equation}}
\renewcommand{\thefigure}{S\arabic{figure}}
\renewcommand{\thetable}{S\arabic{table}}

\section{Common definitions}
The following terms are common to a number of quantum engines and are used throughout this Supplementary Information.
\begin{itemize}[leftmargin=*]
\item The "k-points distance" is a quantity used by almost every implementation to set the k-points mesh for the calculation of integrals in reciprocal space. It indicates the distance between adjacent k-points in the selected mesh. In other words, the number of k-points along each reciprocal axis $i$ is the closest integer to $|b_i|/$"k-points distance" where $b_i$ indicated the reciprocal lattice vector along $i$.
\item The "forces threshold" is the target threshold for the forces during the relaxation process. A relaxation is considered converged when forces on all atoms are below "forces threshold". This is the same quantity referred to as \texttt{threshold\_forces} in the main text of the manuscript.
\item The "stress threshold" is the target threshold for the stress. A relaxation with variable cell is considered converged when the maximum stress component is smaller than "stress threshold". It is the same quantity referred to as \texttt{threshold\_stress} in the main text of the manuscript.
\item The term "PW cutoff" indicates the plane wave cutoff energy, a quantity commonly used to set the size of the basis set in plane-wave \gls{dft} calculations.

\item The $\Delta$ factor quantifies the difference between two \gls{eos} calculations\cite{Lejaeghere:2014}. It has become the standard for the comparison between \gls{dft} methods and codes\cite{Lejaeghere:2016}.

\item The "BFGS algorithm" refers to the Broyden–Fletcher–Goldfarb–Shanno algorithm for solving nonlinear optimization problems. The "L-BFGS alghoritm" is its Limited-memory approximation\cite{Liu:1989}.

\item The "FIRE algorithm" refers to the Fast Inertial Relaxation Engine\cite{Bitzek:2006}.
\end{itemize}

\section{Overview of supported features and completed workflows}
\ctref{tab:matrix-supported-features} provides an overview of which arguments and what values for those arguments are supported by the implementations of the common relax workflow interface for the various quantum engines.

\begin{table*}[!h]
\addtolength{\tabcolsep}{8pt}
\begin{tabular*}{\textwidth}{l l l l}
\toprule
Code      & \texttt{relax\_type} & \texttt{electronic\_type} & \texttt{spin\_type}\\
\midrule
\abinit    & all                                                                         & \texttt{insulator}, \texttt{metal}  & \texttt{`none'}, \texttt{`collinear'}\\
\bigdft    & \texttt{none}, \texttt{positions}                                           & \texttt{insulator}, \texttt{metal}  & \texttt{`none'}, \texttt{`collinear'} \\
\castep    & all                                                                         & (ignored)                           & \texttt{`none'}, \texttt{`collinear'} \\
\cptwok    & \texttt{none}, \texttt{positions}, \texttt{positions\_cell}                 & \texttt{insulator}, \texttt{metal}  & \texttt{`none'}, \texttt{`collinear'}\\
\fleur     & \texttt{none}, \texttt{positions}                                           & (ignored)                           & \texttt{`none'}, \texttt{`collinear'} \\
\gaussian  & \texttt{none}, \texttt{positions}                                           & (ignored)                           & \texttt{`none'}, \texttt{`collinear'} \\
\nwchem    & \texttt{none}, \texttt{positions}, \texttt{positions\_cell}, \texttt{cell}  & \texttt{insulator}, \texttt{metal}  & \texttt{`none'} \\
\orca      & \texttt{none}, \texttt{positions }                                          & (ignored)                           & \texttt{`none'}, \texttt{`collinear'}\\
\qe        & all                                                                         & \texttt{insulator}, \texttt{metal}  & \texttt{`none'}, \texttt{`collinear'}\\
\siesta    & \texttt{none}, \texttt{positions}, \texttt{positions\_shape}, \texttt{cell} & (ignored)                           & \texttt{`none'}, \texttt{`collinear'}\\
\vasp      & all                                                                         & (ignored)                           & \texttt{`none'}, \texttt{`collinear'}\\
\bottomrule
\end{tabular*}
\addtolength{\tabcolsep}{1pt}
\label{tab:matrix-supported-features}
\caption{
    Supported features by the various common relax workflow implementations.
    In addition to the tabulated features, all codes support the three protocols \texttt{`fast'}, \texttt{`moderate'} and \texttt{`precise'}.
}
\end{table*}

\ctref{tab:matrix-completed-workflows} shows which workflows, as described in the main text, have been successfully completed for each quantum engine, with an explanation for those that are missing.

\begin{table*}[!h]
\addtolength{\tabcolsep}{12pt}
\begin{tabular*}{\textwidth}{l c c c c c c c c}
\toprule
\multirow{2}{*}{Code}      & \multicolumn{5}{c}{EOS} & DC & \multicolumn{2}{c}{OPT} \\
          & Si & Al & GeTe & Fe-f & Fe-af & H$_{2}$ & NH$_{3}$-pl & NH$_{3}$-py \\
\midrule
\abinit   & \cmark       & \cmark       & \cmark       & \cmark       & \cmark       & \cmark       & \cmark & \cmark \\
\bigdft   & \cmark       & \cmark       & \nmark$^{a}$ & \xmark$^{c}$ & \xmark$^{c}$ & \xmark$^{c}$ & \cmark & \cmark \\
\castep   & \cmark       & \cmark       & \cmark       & \cmark       & \cmark       & \cmark       & \cmark & \cmark \\
\cptwok   & \cmark       & \cmark       & \nmark$^{a}$ & \xmark$^{d}$ & \xmark$^{d}$ & \cmark       & \cmark & \cmark \\
\fleur    & \cmark       & \cmark       & \nmark$^{a}$ & \cmark       & \cmark       & \xmark$^{e}$ & \cmark & \cmark \\
\gaussian & \nmark$^{f}$ & \nmark$^{f}$ & \nmark$^{f}$ & \nmark$^{f}$ & \nmark$^{f}$ & \cmark       & \cmark & \cmark \\
\nwchem   & \cmark       & \cmark       & \nmark$^{a}$ & \nmark$^{b}$ & \nmark$^{b}$ & \nmark$^{b}$ & \cmark & \cmark \\
\orca     & \nmark$^{g}$ & \nmark$^{g}$ & \nmark$^{g}$ & \nmark$^{g}$ & \nmark$^{g}$ & \cmark       & \cmark & \cmark \\
\qe       & \cmark       & \cmark       & \cmark       & \cmark       & \cmark       & \cmark       & \cmark & \cmark \\
\siesta   & \cmark       & \cmark       & \cmark       & \cmark       & \cmark       & \cmark       & \cmark & \cmark \\
\vasp     & \cmark       & \cmark       & \cmark       & \cmark       & \cmark       & \cmark       & \cmark & \cmark \\
\bottomrule
\end{tabular*}
\label{tab:matrix-completed-workflows}
\caption{
    Overview of which workflow, presented in the main article, could be run by each of the eleven quantum engines.
    A checkmark (\cmark) means that the workflow was successfully completed, a circle (\nmark) means the quantum engine does not support the required functionality or structure and a crossmark (\xmark) means the AiiDA plugin for the quantum engine does not yet support a required feature or it could not successfully run the workflow.
    For the latter two cases, a detailed explanation will be given below in the caption.
    Fe-f and Fe-af stand for ferromagnetic and anti-ferromagnetic iron, respectively.
    \emph{(a)} The code does not support geometry optimizations at constant volume of the cell.
    \emph{(b)} The code does not support running metallic systems which require smeared occupations.
    \emph{(c)} \bigdft{} and its AiiDA plugin both support metallic systems, however, this requires specific mixing inputs which are still experimental.
    Although they give good results for the Al test case, the results are not trustworthy for the Fe examples and are therefore not included in the paper.
    The solution to address the issue is known: a two-steps approach is necessary, restarting the first calculation with a lower electronic temperature.
    This two-steps approach has been successfully used in earlier versions of PyBigDFT to perform the $\Delta$ test, however, the AiiDA plugin and implementation of the common workflow interface for \bigdft{} do not yet support it.
    \emph{(d)} The results for \cptwok{} for the iron structure seemed to converge to an incorrect ground state for as of yet unknown reasons and therefore the results have been omitted.
    \emph{(e)} In certain rare situations, \fleur{} is unable to retain the specified initial magnetization when generating the start density.
    This sometimes leads to calculations wrongly converging to non-magnetic metastable solutions.
    This can be fixed by explicitly breaking the magnetic symmetry by means of a small electric field or enforcing initial occupation values, but neither option is currently supported by the implementation of the common workflow interface for \fleur{}.
    \emph{(f)} The code is not designed for extended systems although there is some support for it, but the AiiDA plugin does not implement it.
    \emph{(g)} The code does not support extended systems.
}
\end{table*}

\section{Code-specific design choices}

We collect here the design choices of every common relax workflow implementation presented in the main text.
These choices include details on the relaxation algorithm, an explanation of the supported features, and the specifications of every protocol implemented.

The textual descriptions in the next sections are meant to provide a quick reference of the most important numerical choices for each code, but cannot be fully complete without becoming cumbersome to read.
This, however, is not an issue: the exact inputs of each run are captured in full detail by AiiDA in the provenance graph~\cite{web:Huber:MCA:2021}.
In addition the implementation of all plugins and workflows is open source and can be inspected to verify the exact choices and rules used to determine the input parameters.

\subsection{\abinit}
\subsubsection{Protocols}
List of \texttt{protocols} supported and their description:
\begin{itemize}[leftmargin=*]
    \item \texttt{fast}.
        This protocol has low precision at minimal computational cost for testing purposes. k-points distance is $0.25$ \AA$^{-1}$. Tolerance on the potential residual is $1\cdot 10^{-7}$ Ha. No additional memory is allowed for basis set enlargement. Forces threshold is $5\cdot 10^{-5}$ Ha/Bohr. Stress threshold is $5\cdot 10^{-3}$ Ha/Bohr$^3$. Fermi-Dirac smearing with broadening 0.008~Ha and 2 times the number of atoms additional bands is set in the case of calculations on metals.
    \item \texttt{moderate}.
        This is the default protocol with normal precision and moderate computational cost. k-points distance is $0.20$ \AA$^{-1}$. Tolerance on the potential residual is $1\cdot 10^{-9}$ Ha. Forces threshold is $5\cdot 10^{-5}$ Ha/Bohr. Stress threshold is $5\cdot 10^{-3}$ Ha/Bohr$^3$. Fermi-Dirac smearing with broadening 0.008~Ha and 2 times the number of atoms additional bands is set in the case of calculations on metals.
    \item \texttt{precise}.
        This protocol should yield fully converged results and is recommended for production calculations that require more precision than provided by the moderate protocol.
        k-points distance is $0.1$ \AA$^{-1}$. Tolerance on the potential residual is $1\cdot 10^{-10}$ Ha. Forces threshold is $5\cdot 10^{-5}$ Ha/Bohr. Stress threshold is $5\cdot 10^{-3}$ Ha/Bohr$^3$. Fermi-Dirac smearing with broadening 0.005~Ha and 2 times the number of atoms additional bands is set in the case of calculations on metals.
\end{itemize}

The PW cutoff is determined automatically thanks to the \texttt{djrepo} table provided by the PseudoDojo initiative~\cite{Setten:2018}.
An additional 15\% memory is booked for the plane wave expansion basis in case of an \texttt{`positions\_cell'} and \texttt{`positions\_volume'} relaxation while only 5\% is used when performing \texttt{`positions\_shape'} for numerical stability.
In all other relaxation types, no additional memory is booked.
In all cases, automatic multilevel parallelization of the calculations is performed directly by the \abinit \- software~\cite{Gonze:2016,Gonze:2020}.
The workflow relies on Projector Augmented-Wave Method (PAW)~\cite{Torrent:2008} pseudopotentials from PseudoDojo~\cite{Setten:2018} in the ``standard'' configuration with PBE exchange-correlation.

\subsubsection{Supported calculation modes}
The \texttt{relax\_type} argument supports all the possible values, meaning that atomic positions and the cell size and shape can be optimized, as well as any combination of those.
All protocols use a L-BFGS algorithm to minimize the forces on the atoms and stress on the cell.

The \texttt{electronic\_type} argument supports the values \texttt{`metal'} and \texttt{`insulator'}. The
\texttt{`insulator'} calculations are performed with fixed occupations, whereas, in \texttt{`metal'} calculations, Fermi-Dirac smearing is employed.

The \texttt{spin\_type} argument supports the values \texttt{`none'} and \texttt{`collinear'}.
In case of \texttt{`collinear'} calculations, unless \texttt{magnetization\_per\_site} is explicitly defined, the maximum theoretical magnetic moment for each element is specified in the z-direction.


The \texttt{reference\_workchain} argument guarantees that the used k-points mesh is identical to the one of the reference workchain.

\subsubsection{Running in the Quantum Mobile virtual machine}
All that is needed to run the \abinit \- common workflows is to install the appropriate PseudoDojo~\cite{Setten:2018} family, which is done as follows:
\begin{lstlisting}[
    label=code:instructions-abinit,
    %caption={
     %   Install the required %PseudoDojo pseudopotential %families for the Abinit common %workflows.
    %}
]
aiida-pseudo install pseudo-dojo -v 1.0 -f jthxml
\end{lstlisting}

\subsection{\bigdft}
\subsubsection{Protocols}
\bigdft{} input generator switches from cubic scaling computation to linear scaling computation if the number of atoms in the system is more than 200. This value is meant to be adapted more finely in future releases to provide a good heuristic.

For cubic computations, inputs are generated by the input generator.
\begin{itemize}[leftmargin=*]
    \item \texttt{fast}.
        This protocol is meant for testing and should provide fast and less accurate results. hgrids is set to 0.45 bohrs and k-points are generated using a real space equivalent length of 20 bohrs.
    \item \texttt{moderate}.
        This protocol the main one for providing reasonable accuracy while staying fast enough. Main difference with fast is that hgrids is set to 0.3, while default k-points distance is set to 40 bohrs (real space). Convergence criteria are also stricter.
    \item \texttt{precise}.
        This protocol aims at providing the most accurate results without taking costs into account. hgrids is set to 0.15. k-points are computed by PyBigDFT according to the number of atoms for this precision level.
\end{itemize}

For linear computations, \bigdft{} provides "linear\_fast", "linear\_moderate", and "linear\_accurate" input sets, which are mapped to \texttt{`fast'}, \texttt{`moderate'}, and \texttt{`precise'} protocols respectively. They use 0.3 hgrids settings with more relaxed or strict convergence criterion.

In order to have more accurate results, it has been decided to override, when possible, the default pseudopotentials used in \bigdft{} with the soft and norm-conserving HGH pseudopotentials (including nonlinear core-correction) generated by S.Saha \cite{Saha:2017}. These pseudopotentials are not yet available in the UPF format and, for the moment, are included directly in PyBigDFT (in PSP8 format).
In the near future, these pseudopotentials will be converted in UPF format and made available to AiiDA users in the standard way (UpfData class and corresponding pseudopotential families). \bigdft{} has to be compiled with PSPIO in order to support UPF pseudopotentials. The version of \bigdft{} in the Quantum Mobile already includes this support.

\subsubsection{Supported calculation modes}
As of now, the \bigdft{} implementation only supports the relaxation of the atomic coordinates. Therefore it handles the \texttt{threshold\_forces} argument, but not the \texttt{threshold\_stress} one. The default applied algorithm for relaxation in the \bigdft{} plugin is FIRE. Cell relaxation is not supported internally in \bigdft{}, but we foresee the creation of a future AiiDA workflow that drives externally the cell relaxation.

IMPORTANT NOTE: \bigdft{} does not yet handle non-orthorhombic cells, a transformation of the input cell is attempted if an invalid cell is provided in input. The computation is then performed on the resulting orthorhombic super-cell. The Si test-case used for in this paper has to be transformed this way before computation. The next release of \bigdft{} will overcome these limitations.

The \texttt{spin\_type} argument supports the values \texttt{`none'} and \texttt{`collinear'}. If the \texttt{`collinear'} option is provided, the computation is launched with spin-polarized enabled.
By default, the plugin will initialize spin through a round-robin scheme over the atoms to match the computed polarity moment. This can be overridden by using the \texttt{magnetization\_per\_site} argument, to provide the values for each atom directly.

The \texttt{electronic\_type} argument supports both \texttt{`metal'} and \texttt{`insulator'}. If metal is specified, specific mixing inputs are added to account for the nature of the system. These inputs are still experimental and they give good results for Al but not for the Fe test case, that is therefore not included in the paper. The solution to address the issue is known: a two-steps approach is necessary, restarting the first calculation with a lower electronic temperature. This two-steps approach has been successfully used in earlier versions of PyBigDFT to perform the $\Delta$ test, however it is not yet included in the implementation of the present project. A future internal workflow will automatically perform the two steps allowing \bigdft{} to return correct results for Fe through the common interface.

\subsubsection{Running in the Quantum Mobile virtual machine}
\bigdft{} is provided in latest Quantum mobile releases. There is no specific configuration to run it, once set up with aiida.

\subsection{\castep}

\subsubsection{Protocols}

List of \texttt{protocols} supported and their description:

\begin{itemize}[leftmargin=*]

    \item \texttt{fast}. This protocol is intended for coarse calculations and quick tests. Many parameters are downgraded from the \texttt{moderate} settings. The main difference is that a different on-the-fly generated (OTFG) pseudopotential library \texttt{QC5} is used, which is designed to have converged  results using a PW cutoff $\approx$ 300 eV for most elements at the cost of transferability. The \texttt{medium} basis set precision settings is applied, with a reduced electronic energy tolerance $1\cdot 10^{-5}$ eV per atom, and a increased k-point distance of 0.25 \AA$^{-1}$ (equivalent to 0.03979 2$\pi$\AA$^{-1}$ in \castep's convention).

    \item \texttt{moderate}. This protocol is intended for general use and expected to give physically sensible results. The k-point distance is set to 0.15 \AA$^{-1}$. Note that \castep \- uses a convention without the explicit $2\pi$ factor when converting between the real and reciprocal space distances, so this setting is equivalent to a \texttt{kpoints\_mp\_spacing} of 0.02387 2$\pi$\AA$^{-1}$.
    The \texttt{fine} basis precision setting is used. With this option, \castep \- internally chooses the PW cutoff based on the convergence data of the core-corrected ultrasoft pseudopotentials to be used, which are on-the-fly generated during the calculation using the built-in \texttt{C19} library of generation settings.
    If the calculation only involves elements for which the pseudopotentials are very soft, such as silicon, the default low PW cutoff can potentially cause problems during variable cell geometry optimisation.
    To avoid this issue, a minimum PW cutoff of 326 eV is imposed in the protocol. The electronic energy convergence tolerance is set to $10^{-6}$ eV per atom with the default tolerance window of three steps.
    The geometry optimisation uses a forces threshold of 0.05 eV/\AA{} in conjunction with a energy tolerance of $2 \cdot 10^-5$ eV per atom, a stress threshold of 0.1 GPa (if applicable), and a atomic position change tolerance of 0.001 \AA, with a convergence window of two ionic steps.

    \item \texttt{precise}. This protocol is intended for high accuracy calculations and obtaining converged results.
    Most parameters are based on the \texttt{moderate} settings, with the \texttt{basis\_precision} setting increased to \texttt{precise}. The electronic energy tolerance is reduced to $1 \cdot 10^{-8}$ eV per atom. The geometry optimisation thresholds are reduced to 0.03 eV/\AA{} for forces, 0.05 GPa for stress and $10 ^{-5}$ eV per atom for the total energy.
    The \texttt{grid\_scale} setting that controlling the FFT grid density is set to 2, and the \texttt{fine\_grid\_scale} is increased to 3.
    This is intended for minimising any FFT aliasing errors. The k-points spacing is further reduced to 0.1 \AA $^{-1}$ (equivalent to 0.01591 2$\pi$\AA $^{-1}$ in \castep's convention) for a improved sampling of the reciprocal space.

\end{itemize}

\subsubsection{Supported calculation modes}

The \texttt{relax\_type} argument supports all the agreed values.
All protocol defaults to the L-BFGS algorithm with line search for performing geometry optimisation.
The only exception is when the \texttt{relax\_type} is \texttt{`positions\_shape'}. In this case the two-point steepest descent optimiser (TPSD) is selected as it
gives superior performance as of \castep{} version 19.1.1.

The \texttt{electronic\_type} argument supports both \texttt{`metal'} and \texttt{`insulator'}, although the value is internally ignored, and density mixing solver is used in both cases with a Gaussian smearing of 0.2 eV applied to the occupations of the electronic bands.

The \texttt{spin\_type} argument supports values \texttt{`none'} and \texttt{`collinear'}. The initial per-site magnetization can be passed to bias the electronic solver into specific spin arrangements. If no per-site magnetization value is passed, \castep{} defaults to have zero magnetisation per site and prints a warning message in the output file. Since it is unlikely that the spin symmetry can be be broken spontaneously, it is recommended that the user always pass the per-site initial magnetisation explicitly.
While \castep{} itself supports non-collinear and spin-orbit calculations, they are not yet enabled through the common interface presented in this work. This is because the common interface \texttt{magnetization\_per\_site} input does not allow yet non-collinear initialization of the spins and, moreover, specially generated norm-conserving potentials are needed for spin-orbit calculations.

The \texttt{reference\_workchain} argument makes sure that the k-points mesh that is used is the same as that used in the defined workchain for
consistent energy comparisons.

\subsubsection{Running in the Quantum Mobile virtual machine}

The source code of CASTEP can be compiled using the following commands on Quantum Mobile.
We assume that the \texttt{aiida-castep} plugin has been installed as one of the dependencies of \texttt{aiida-common-workflows}.

If internet connection is available, the following command can be used to install OpenBLAS in Quantum Mobile:

 \begin{lstlisting}[
 label=code:library-openblas,
 numbers=left,
 caption={
 Commands to install the OpenBLAS library.
 }
 ]
 sudo apt-get update
 sudo apt-get install libopenblas-dev
 \end{lstlisting}

The availability of OpenBLAS may improve the execution speed of the code, but it is not essential.
It is also possible to link with intel MKL library, but it is beyond the scope of this guide.

Assuming a source archive of CASTEP version 19.1.1 is placed in the working directory, paste the following commands into the terminal to compile a binary and setup the \texttt{Code} node for AiiDA:

\begin{lstlisting}[
label=code:instructions-castep,
numbers=left,
caption={
Compiling a CASTEP executable on Quantum Mobile and settings up the \texttt{Code} node.
}
]
cat > install-castep.sh << EOF
set -e
set -x
tar zxvf CASTEP-19.11.tar.gz
cd CASTEP-19.11
apt list --installed | grep libopenblas-dev > /dev/null
if [ \$? -eq 0 ]; then
mathlib=openblas
else
mathlib=default
fi
set +e
make MATHLIB=\$mathlib FFT=fftw3 COMMS_ARCH=mpi SUBARCH=mpi TARGETCPU=portable -j `nproc` clean
set -e
make MATHLIB=\$mathlib FFT=fftw3 COMMS_ARCH=mpi SUBARCH=mpi TARGETCPU=portable -j `nproc`
make MATHLIB=\$mathlib FFT=fftw3 COMMS_ARCH=mpi SUBARCH=mpi TARGETCPU=portable -j `nproc` install
EOF
bash install-castep.sh

cat > castep-code.yaml << EOF
label: castep-19.1.1
description: CASTEP 19.1.1 Compiled locally
input_plugin: castep.castep
on_computer: True
computer: localhost
remote_abs_path: /home/max/CASTEP-19.11/bin/bin/linux_x86_64_gfortran7.0--mpi/castep.mpi
prepend_text: 'export OMP_NUM_THREADS=1'
append_text: ''
EOF
workon aiida
verdi code setup --config castep-code.yaml

\end{lstlisting}

During the compilation, the user will be prompted to confirm the path to the libraries, and directory where the compiled binary to be installed.
In both cases, press the Enter to use the default option.

Free-of-charge source code licenses for \castep\ can be obtained for academic use. This option is available to  the researchers world-wide. More details can be found one the \href{http://www.castep.org/}{official  website}.

\subsection{\cptwok}

\subsubsection{Protocols}
Every protocol employs Gaussian and plane waves\cite{Vandevondele:2005} method to compute energy and forces. A multi-grid approach is used to represent the electron density and the product Gaussian functions. Several parameters are required to define the multi-grid. CUTOFF (Ry) parameter is the plane wave cutoff of the finest grid used to map the electron density. REL\_CUTOFF  (Ry) determines how Gaussians are mapped into the multi-grid. NGRIDS parameter defines the total number of grids. EPS\_DEFAULT acts as a default value for many other parameters trying to achieve the precision in energy up to the value of EPS\_DEFAULT.  All protocols have a default maximum force threshold of $4.5\cdot 10^{-4}$ Hartree per {\bohr} and root mean square force threshold of  $3.0\cdot 10^{-4}$ Hartree per {\bohr}.

Here is the list of supported \texttt{protocols} and the values of the parameters:

\begin{itemize}[leftmargin=*]
    \item \texttt{fast}. This protocol is intended for testing purposes and uses very loose settings. The k-points distance is set to $1.0$ \AA$^{-1}$. The multi-grid parameters are: 400 Ry CUTOFF, 30 Ry REL\_CUTOFF, and 4 NGRIDS. Target accuracy for the SCF convergence is set to  $1 \cdot 10^{-6}$ and EPS\_DEFAULT is set to $1 \cdot 10^{-10}$.

    \item \texttt{moderate}. This protocol is intended for general use and is expected to provide sensible results. The k-points distance is set to $0.5$ \AA$^{-1}$. The multi-grid parameters are: 600 Ry CUTOFF, 40 Ry REL\_CUTOFF, and 4 NGRIDS. Target accuracy for the SCF convergence is set to  $1 \cdot 10^{-7}$ and EPS\_DEFAULT is set to $1 \cdot 10^{-12}$.

    \item \texttt{precise}. This protocol is intended for high-accuracy calculations. The k-points distance is set to $0.1$ \AA$^{-1}$. The multi-grid parameters are: 1000 Ry CUTOFF, 50 Ry REL\_CUTOFF, and 3 NGRIDS. Target accuracy for the SCF convergence is set to  $1 \cdot 10^{-8}$ and EPS\_DEFAULT is set to $1 \cdot 10^{-16}$.
\end{itemize}

\subsubsection{Supported relax types and calculation modes}
The list of currently supported \texttt{relax\_type} is: \texttt{`none'}, \texttt{`positions'}, \texttt{`positions\_cell'}. All protocols use the standard BFGS algorithm to optimize the atomic positions and the cell.

The \texttt{electronic\_type} argument supports the values \texttt{`metal'} and \texttt{`insulator'}. For \texttt{`insulator'}, the calculation is performed using the orbital transformation method with fixed occupations. The default total magnetic moment is equal to $0$ for an even number and $1$ for an odd number of electrons. For \texttt{`metal'} a diagonalisation with Fermi-Dirac smearing is used at electronic temperature of $500$ K and $20$ molecular orbitals are added for each spin. In case smearing is employed, the total magnetization is flexible.

The \texttt{spin\_type} argument supports the values \texttt{`none'} and \texttt{`collinear'}. In the case of the \texttt{`collinear'} option, the spin-polarized calculation are enabled. Additionally, the user can specify the \texttt{magnetization\_per\_site}, which would be ignored in the \texttt{`none'} case. If the user provides the \texttt{magnetization\_per\_site}, the total multiplicity is derived from it.

\subsubsection{Running in the Quantum Mobile virtual machine}
No additional setup is required to run \cptwok{} on Quantum Mobile: the code comes already preinstalled and the necessary data files are shipped together with the \texttt{aiida-common-workflows} package.

\subsection{\fleur}
The execution of the \fleur{} implementation of the common relax workflows relies on the \texttt{FleurBaseRelaxWorkChain}, \texttt{FleurRelaxWorkChain}, \texttt{FleurScfWorkChain} and \texttt{FleurBaseWorkChain} of the \texttt{aiida-fleur} package \cite{Broeder:2019}.
\subsubsection{Protocols}
The \fleur{} program sets most internal parameters during generation of the full input to appropriate values. The protocols for \fleur{} contain two kinds of values, the first influences the behavior of the underlying workflows, while the second explicitly sets some internal parameters during input generation. All protocols have a default force threshold of $1\cdot 10^{-3}$ Hartree per \bohr, which can be overwritten by user through the dedicated input of the relax common workflow.
List of \texttt{protocols} supported and their description:
\begin{itemize}[leftmargin=*]
    \item \texttt{fast}. In this protocol parameters are set in a way to reduce the computational cost. The protocol should still yield reasonable results but should not be used for production calculations, merely for testing. The most important parameters are: k-point distance is set to $0.4$~\AA$^{-1}$, self-consistency convergence threshold for charge density to $2\cdot 10^{-7}$ electrons/\bohr$^{3}$, k-max basis cut-off to $3.2$ \bohr$^{-1}$, maximal number of relaxation calls to three, maximal number of iterations per SCF workflow to 240.
    \item \texttt{moderate}. This protocol provides a reasonable trade-off between accuracy and the computational cost. The most important parameters are: k-point distance is set to $0.2$ \AA$^{-1}$, self-consistency convergence threshold for charge density to $2\cdot 10^{-8}$ electron/\bohr$^{3}$, k-max basis cut-off to $4.0$ \bohr$^{-1}$, maximal number of relaxation calls to five, maximal number of iterations per SCF workflow to 240.
    \item \texttt{precise}. This protocol is based on the moderate protocol but various parameters have been changed to improve the precision at the expenses of an increased computational cost. This protocol should yield fully converged results and is recommended for production calculations that require higher precision than provided by the moderate one. The calculation parameters are: k-point distance is set to $0.1$ \AA$^{-1}$, self-consistency convergence threshold for charge density to $2\cdot 10^{-9}$ electron/\bohr$^{3}$, k-max basis cut-off to $5.0$ \bohr$^{-1}$, maximal number of relaxation calls to ten, maximal number of iterations per SCF workflow to 360.
\end{itemize}

For all other parameters within the Full-potential Linearized-Augmented-Plane-Wave method (FLAPW) we rely on the default choices of \fleur{} and the \fleur{} input generator. Consistency of these is only enforced within workflows where total energies need to be compared through the \texttt{reference\_workchain} argument. Regarding the choice of k-point mesh, the use of a single k-point along directions having no periodic conditions is enforced. For example, a film calculation will always have a single k-point along z-direction.

\subsubsection{Supported calculation modes}
The \texttt{relax\_type} argument supports the \texttt{`positions'} and \texttt{`none'} values, meaning that atomic positions can be optimized. All protocols use the standard BFGS algorithm implemented in \fleur{} to minimize forces acting on atoms. However, first several iterations will be run using the straight mixing and BFGS is used only after the largest force is small enough. This is done to suppress the common problem of muffin-tin-overlap because BFGS tends to propose too large displacements during the first steps, moving atoms far away from the equilibrium positions and hence overlapping them.

The \texttt{electronic\_type} argument supports the values \texttt{`metal'} and \texttt{`insulator'}. Simulations for \texttt{`insulator'} and \texttt{`metal'} are treated in the current protocol in the same way. In general, for calculation of metals with \fleur{} it might be necessary to use a denser k-point mesh or introduce a larger smearing around the Fermi energy. The k-point distance of $0.15$ \AA$^{-1}$ for the precise protocol is expected to be a reasonable compromise to treat both \texttt{electronic\_type}s in the same way.

The \texttt{spin\_type} argument supports the values \texttt{`none'} and \texttt{`collinear'}.
In case of \texttt{`collinear'} calculations, unless \texttt{magnetization\_per\_site} is explicitly defined, a non-zero starting magnetization is chosen where the value for each site is determined by the input generator of \fleur{}. If \texttt{magnetization\_per\_site} is defined, starting magnetic moments for given sites are set in the input generator, which might also require explicitly breaking the symmetry of the crystal structure.

The \texttt{reference\_workchain} argument guarantees that the FLAPW parameters and species parameters, including basis set cutoffs, muffin-tin radii are consistent to allow high accuracy energy differences between these simulations.

\subsubsection{Running in the Quantum Mobile virtual machine}
Since everything needed to run the \fleur{} common workflows is already installed on Quantum Mobile, no further steps are needed.

\subsection{\gaussian}

The Gaussian workflows have been tested and all the results shown have been calculated using the Gaussian 09 Revision D.01 \cite{Frisch:2016} but the presented implementation should work with most versions of Gaussian, as the input and output syntax of the demonstrated calculations has remained the same.

\subsubsection{Protocols}
\gaussian{} has many internal checks to automatically set most computational parameters at appropriate values. The protocol specification tries to take advantage of this by only varying the basis set size, integration grid and optimization tolerance. Additionally, symmetry is disabled (route parameter \texttt{NoSymm}) in all cases to make the protocols more generally applicable. List of supported \texttt{protocols} and their descriptions:
\begin{itemize}[leftmargin=*]
    \item \texttt{fast}. A fast protocol that is mainly used for testing, it uses a small \texttt{Def2SVP\cite{Weigend:2005}} basis set and a loose (\texttt{opt=loose}) geometry optimization tolerance.
    \item \texttt{moderate}. A protocol with moderate accuracy, using the \texttt{Def2TZVP\cite{Weigend:2005}} basis set, \texttt{ultrafine} integration grid and the default geometry optimization tolerance.
    \item \texttt{precise}. A protocol using the \texttt{Def2QZVP\cite{Weigend:2003}} basis set, \texttt{superfine} integration grid and tight (\texttt{opt=tight}) optimization tolerance.
\end{itemize}

\subsubsection{Supported relax types and calculation modes}
The supported \texttt{relax\_type} values are \texttt{`none'}, which just performs a force calculation without any structural optimization (\gaussian{} keyword \texttt{force}), and \texttt{`positions'}, which uses the standard \gaussian{} optimization algorithm to optimize the atomic positions.

The \texttt{electronic\_type} input is ignored (\texttt{`metal'} and \texttt{`insulator'} are treated the same way).

The \texttt{spin\_type} \texttt{`none'} and \texttt{`collinear'} are supported and correspond to a restricted (RKS) and unrestricted Kohn-Sham (UKS) calculation, respectively.
In case of \texttt{`collinear'} calculations, if no \texttt{magnetization\_per\_site} is explicitly passed, the lowest allowed spin multiplicity is specified (1 in case of even and 2 in case of odd electrons).
If \texttt{magnetization\_per\_site} is passed, the site-specific spin information is disregarded and only total spin guess (sum of \texttt{magnetization\_per\_site}) is processed to determine the input spin multiplicity (rounded to the closest allowed spin multiplicity).
In case of a calculation of the open-shell singlet, HOMO and LUMO are mixed to break the spin-symmetry (\gaussian{} route parameter \texttt{guess=mix}).

The \texttt{reference\_workchain} argument is ignored.

\subsubsection{Running in the Quantum Mobile virtual machine}

After the user sets up their \gaussian{} code in the standard AiiDA manner, the common workflows are available without any further setup.

\subsection{\nwchem}
\subsubsection{Protocols}

List of \texttt{protocols} supported and their description:
\begin{itemize}[leftmargin=*]
    \item \texttt{fast}. The minimum k-point distance is set to $1.0$ \AA$^{-1}$ and an SCF energy convergence tolerance of $1.0e^{-5}$ Ha is set. The geometry convergence is set to \texttt{loose}, corresponding to a forces threshold on any atom of $4.5e^{-3}$ Ha bohr$^{-1}$, and a root mean square (RMS) gradient of $3.0e^{-3}$ Ha bohr${^-1}$
    \item \texttt{moderate}. The minimum k-point distance is set to $0.2$ \AA$^{-1}$ and an SCF energy convergence tolerance of $1.0e^{-7}$ Ha is set. The geometry tolerance is set to \texttt{default}, corresponding to a forces threshold on any atom of $4.5e^{-4}$ Ha bohr$^{-1}$, and an RMS gradient of $3.0e^{-4}$ Ha bohr${^-1}$
    \item \texttt{precise}. The minimum k-point distance is set to $0.1$ \AA$^{-1}$ and an SCF energy convergence tolerance of $1.0e^{-9}$ Ha is set. The geometry tolerance is set to \texttt{tight}, corresponding to a forces threshold on any atom of $1.5e^{-5}$ Ha bohr$^{-1}$, and an RMS gradient of $1.0e^{-5}$ Ha bohr${^-1}$
\end{itemize}

For all protocols, the PBE exchange-correlation functional is used with Hamann and Troullier-Martins norm-conserving psuedopotentials.
In the case of molecular calculations run using the common interface, the plane wave code is also used rather than the main DFT module in \nwchem{} which employs localised basis sets.
This for consistency between protocols. The main differences are that the gamma-point only code is used, and that a larger cutoff of 140 Ha is set for all protocols.

\subsubsection{Supported relax types and calculation modes}

The \texttt{relax\_type} supported are: \texttt{`none'}, \texttt{`positions'}, \texttt{`positions\_cell'}, and \texttt{`cell'}.
The relaxation of the atomic coordinates and cell vectors is performed using the \texttt{DRIVER} module, which implements a quasi-newton optimization with line searches and approximate energy Hessian updates.

If \texttt{electronic\_type} is set as \texttt{`insulator'}, the wavefunction is optimized using Grassmann L-BFGS algorithm. If smearing is required, by setting \texttt{electronic\_type} to \texttt{`metal'}, a band-by-band optimiser is used with Fermi-Dirac smearing.

Spin polarized calculations are not supported yet through the common interface, they will be enabled in the near future.

\subsubsection{Running in the Quantum Mobile virtual machine}

Other than adding the \nwchem{} code in the standard AiiDA manner, the common workflows are available without any further setup.

\subsection{\orca}
\subsubsection{Protocols}
The following protocols are defined to be used to a wide range of systems considering the required accuracy/cost by the user. These protocols provide higher accuracy via increasing the size of the basis set and tightening the convergence criteria for self-consistent field and geometry convergence. We used recent versions of Ahlrichs basis sets\cite{Weigend:2003,Weigend:2005} in the presented protocols. It is noteworthy that the latter criteria can be easily altered via invoking proper keywords.
\begin{itemize}[leftmargin=*]
    \item \texttt{fast}. It uses Def2-SVP\cite{Weigend:2005} basis along with scf and geometry optimization convergence criteria of Strong and LOOSOPT, respectively, to provide quick setup for testing purposes.
    \item \texttt{moderate}. We used triple zeta Def2-TZVP\cite{Weigend:2005} basis set and decreased the electronic and geometry optimization to Tight and NORMALOPT which provides a fair level of accuracy.
    \item \texttt{precise}. Increased basis set size to quadruple zeta (Def2-QZVP\cite{Weigend:2003}) along with tighter scf and geometry tolerance of VeryTight and TIGHTOPT, respectively.
\end{itemize}

\subsubsection{Supported relax types and calculation modes}
The \orca{} implementation supports the \texttt{`none'} and \texttt{`positions'} values for the \texttt{relax\_type}. The former one is intended for single point calculation and the latter one for the geometry optimization.

\texttt{`metal'} and \texttt{`insulator'} have the same effect as \texttt{electronic\_type} input; both will be ignored.

Restricted and unrestricted Kohn-Sham calculations can be requested by setting the  \texttt{spin\_type} to \texttt{`none'} and \texttt{`collinear'}, respectively. In the latter case if \texttt{magnetization\_per\_site} is NOT provided the spin multiplicity of 1 and 2 will be set for even and odd number of electrons in the system, respectively, if \texttt{magnetization\_per\_site} is given, the site-specific spin information is ignored and only total spin guess (sum of \texttt{magnetization\_per\_site}) is processed to determine the input spin multiplicity (rounded to the closest allowed spin multiplicity).

This implementation does not use the \texttt{reference\_workchain}. Also, as \orca{} internally sets thresholds for forces based on the selected geometry optimization convergence criteria, we do not explicitly define them.

\subsubsection{Running in the Quantum Mobile virtual machine}

\orca{} is a free-ware for academic users for academic usage and can be obtained via registering on the official website (https://orcaforum.kofo.mpg.de). All other uses can request a license at www.faccts.de. \orca{} comes as two versions with static and shared libraries and requires an MPI engine that can be either OpenMPI or MPICH. The shared version is recommended to be used within Quantum Mobile as it requires less disk space. Afterwards, \orca{} can be setup following the instructions provided in detail in AiiDA documentations and be used for running \orca{} implementation of the present work. It should be noted that in case of using shared version of \orca{}, \texttt{LD\_LIBRARY\_PATH} environmental variable should be set to the location of \orca{} in Quantum Mobile.

\subsection{\qe}
\subsubsection{Protocols}

All of the \qe{}\cite{Giannozzi:2009,Giannozzi:2017} protocols use the \gls{sssp}\cite{Prandini:2018} v1.1.
Below is a list the supported \texttt{protocols} and their description:
\begin{itemize}[leftmargin=*]
    \item \texttt{fast}.
        The fast protocol is designed to yield reasonable results at minimal computational cost and should only be used for testing and demonstration purposes.
        It uses the efficiency configuration of the \gls{sssp} and the following precision settings: k-points distance is $0.5$ \AA$^{-1}$, self-consistency convergence threshold is $0.4\cdot 10^{-9}$ Ry per atom, energy threshold for the ionic convergence is $1\cdot 10^{-4}$ Ry per atom and forces threshold is $1\cdot 10^{-3}$ Ry/bohr.
        The convergence on the stress is left to the default of the code which is $0.05$ GPa.
    \item \texttt{moderate}.
        The moderate protocol is the default protocol used for production calculations.
        It uses the efficiency configuration of the \gls{sssp} and the following precision settings: k-points distance of $0.15$ \AA$^{-1}$, self-consistency convergence threshold of $0.2\cdot 10^{-9}$ Ry per atom, energy threshold for the ionic convergence of $1\cdot 10^{-5}$ Ry per atom and forces threshold of $1\cdot 10^{-4}$ Ry/bohr.
        The convergence on the stress is left to the default of the code which is $0.05$ GPa.
    \item \texttt{precise}.
        The precise protocol should yield fully converged results and is recommended for production calculations that require more precision than provided by the moderate protocol.
        It uses the precisison configuration of the \gls{sssp} and the following precision settings: k-points distance is $0.1$ \AA$^{-1}$, self-consistency convergence threshold is $0.1\cdot 10^{-9}$ Ry per atom, energy threshold for the ionic convergence is $0.5\cdot 10^{-5}$ Ry per atom and forces threshold is $0.5\cdot 10^{-4}$ Ry/bohr.
        The convergence on the stress is left to the default of the code which is $0.05$ GPa.
\end{itemize}

The protocol settings listed above are the result of a rigorous study that will be detailed in a forthcoming publication.

\subsubsection{Supported calculation modes}
The \texttt{relax\_type} argument supports all the agreed values, meaning that atomic positions and the cell size and shape can be optimized, as well as any combination of those.
All protocols will use the standard BFGS algorithm to minimize the forces on the atoms and stress on the cell.

The \texttt{electronic\_type} argument supports the values \texttt{`metal'} and \texttt{`insulator'}.
For \texttt{`insulator'}, the calculation is performed with fixed occupations, whereas for \texttt{`metal'} a Marzari-Vanderbilt smearing \cite{Marzari:1999} is employed with a broadening of $0.01$ Ry.

The \texttt{spin\_type} argument supports the values \texttt{`none'} and \texttt{`collinear'}.
In case of \texttt{`collinear'} calculations, unless \texttt{magnetization\_per\_site} is explicitly defined, a non-zero starting magnetization is chosen where the value for each site is element dependent.

The \texttt{reference\_workchain} argument guarantees that the k-points mesh that is used is identical to that used in the defined workchain.

\subsubsection{Running in the Quantum Mobile virtual machine}
All that is needed to run the \qe{} common workflows is to install the \gls{sssp} families, which is done as follows:
\begin{lstlisting}[
    label=code:instructions-qe,
    numbers=left,
    caption={
        Install the required SSSP pseudopotential families for the \qe{} common workflows.
    }
]
aiida-pseudo install sssp -p efficiency
aiida-pseudo install sssp -p precision
\end{lstlisting}

\subsection{\siesta}
The siesta implementation of the common relax workflow can be used only with a version of the \siesta \- code that support the use of pseudopotentials in "psml" format. The \href{https://gitlab.com/siesta-project/siesta/-/wikis/Guide-to-Siesta-versions}{wiki page} of the \siesta{} code can be checked for up-to-date information on the compatble versions.

\subsubsection{Protocols}
All the \siesta \- protocols use the same set of pseudopotentials: the "standard" scalar relativistic set of PseudoDojo\cite{Setten:2018}, which are norm-conserving pseudopotentials.
We warn users that the optimal basis set for a \siesta{} calculation depends on the chemical environment, therefore an optimization procedure should be carried on before the study of any new system.
The statically chosen basis for each chemical species included in the protocols must be considered a good starting point, but not assumed to be optimal in any possible environment. In fact, they have been tested only on crystal elements through the $\Delta$ test. More about the tests of this protocols will be detailed in a forthcoming publication.

The list of supported \texttt{protocol}s and their description follows.
\begin{itemize}[leftmargin=*]
    \item \texttt{fast}. Protocol running \siesta{} with standard inputs for testing. This includes an automatically generated basis ``DZ'' (Double-Zeta, two radial functions per angular momentum channel) with \texttt{energy-shift} of 200 meV, a \texttt{mesh-cutoff} of 50 Ry and a k-points distance of 0.2~$\text{\AA}^{-1}$. Tolerance for the density matrix is set to $10^{-3}$. The forces and stress thresholds are 0.04 eV/Ang and 1 GPa respectively.
    \item \texttt{moderate}. A protocol combining simple basis selection, moderate computational resources and  satisfactory results in the comparison with all-electron references.
    It globally sets basis ``DZP'' (Double-Zeta-Polarized, two radial functions per angular momentum channel plus an orbital obtained from the polarization of the highest occupied orbital)  with \texttt{energy-shift} of 50 meV, a \texttt{mesh-cutoff} of 200 Ry and a k-points distance of 0.1~$\text{\AA}^{-1}$. However custom basis (manually selected basis orbitals radii) are employed for the Ca, Sr and Ba elements, since the DZP choice resulted in too-large radius for the “s” orbitals. Tolerance for the density matrix is set to $10^{-4}$. The forces and stress thresholds are 0.04 eV/Ang and 1 GPa respectively.
    \item \texttt{precise}. Protocol with stringent settings and optimized basis for crystal elements.
    It globally sets a \texttt{mesh-cutoff} of 500 Ry and a k-points distance of 0.1~$\text{\AA}^{-1}$.
    Basis sizes and orbital radii are customized for each chemical species through an optimization procedure based on the basis enthalpy minimization for crystal elements. Tests based on the $\Delta$ test confirm the effectiveness of the choice.
    Tolerance for the density matrix is set to $10^{-5}$. The forces and stress thresholds are 0.005 eV/Ang, 0.7 GPa respectively.
\end{itemize}

\subsubsection{Supported relax types and calculation modes}
The \texttt{relax\_type} supported are: \texttt{`none'}, \texttt{`positions'}, \texttt{`positions\_shape'} and \texttt{`cell'}.
The relaxation of atomic coordinates is performed with the conjugate-gradient algorithm until reaching the forces threshold value.
In the case of variable-cell relaxation, the minimization targets zero hydrostatic pressure through the conjugate-gradient algorithm.

The \texttt{electronic\_type} \texttt{`metal'} and \texttt{`insulator'} are treated in the same way. \siesta{} calculations on metallic systems usually require a denser k-points grid compared to insulators, however, at least for the \texttt{precise} and \texttt{moderate} protocols, the selected k-points distance is expected to generate a sufficiently dense mesh to treat both \texttt{electronic\_types} in the same way.

The \texttt{spin\_type} \texttt{`none'} and \texttt{`collinear'} are supported.
In case of \texttt{`collinear'} calculations, if no \texttt{magnetization\_per\_site} is explicitly passed, a ferromagnetic arrangement is imposed as initial magnetization, with maximum atomic moment on each atom.

The \texttt{reference\_workchain} argument generates a calculation with the same k-points mesh and same real space mesh of the  \texttt{reference\_workchain}. The real space mesh is set using the \siesta{} keyword \texttt{mesh-sizes}.

\subsubsection{Running in the Quantum Mobile virtual machine}

In the Quantum Mobile, the creation of a pseudopotential family with name \texttt{nc-sr-04\_pbe\_standard\_psml} is necessary in order to run \siesta{} calculations through the common interface.
This is achieved simply typing in the command line:
\begin{lstlisting}[
    label=code:siestapseudo,
    numbers=left,
    caption={
        Command to set up the pseudopotential family required by the \siesta{} implementation of the relax common workflow.
    }
]
verdi data psml uploadfamily /usr/local/share/siesta/psml-files-qm/nc-sr-04_pbe_standard/ nc-sr-04_pbe_standard_psml "pseudos from PseudoDojo"
\end{lstlisting}

\subsection{\vasp}
All results in this work have been produced with \vasp{} 5.4.4 and AiiDA-\vasp{} 2.1.0. The potentials used are based on the standard PBE potentials supplied with the same \vasp{} version. The precise protocol was used to generate all results. The Projector Augmented-Wave method (PAW) was used\cite{Kresse:1999}.
\subsubsection{Protocols}
List of \texttt{protocols} supported and their differences:
\begin{itemize}[leftmargin=*]
    \item \texttt{fast}. Low precision, minimal computational cost. For testing purposes. k-points distance is $0.25$ \AA$^{-1}$. Maximum forces threshold is $1.0 \times 10^{-1}$ eV/\AA. \texttt{PREC} is \texttt{Single}. \texttt{EDIFF} is $1.0 \times 10^{-4}$. The protocol imposes at least six electronic steps per self-consistent cycle. A conjugate gradient algorithm for relaxing the positions is implemented.
    \item \texttt{moderate}. Standard precision, moderate computational cost. k-points distance is $0.15$ \AA$^{-1}$.
    Maximum forces threshold is $1.0 \times 10^{-2}$ eV/\AA. \texttt{PREC} is \texttt{Normal}. \texttt{EDIFF} is $1.0 \times 10^{-5}$. The protocol imposes at least four electronic steps per self-consistent cycle. A quasi-Newton algorithm for relaxing the positions.
    \item \texttt{precise}. Elevated precision, high computational cost. k-points distance is $0.10$ \AA$^{-1}$. Maximum forces threshold is $1.0 \times 10^{-3}$ eV/\AA. \texttt{EDIFF} is $1.0 \times 10^{-6}$. \texttt{PREC} is \texttt{Accurate}. A quasi-Newton algorithm for relaxing the positions.
\end{itemize}

For all the workflows, the following was used: Gaussian smearing with width of 0.2 eV for the integration and a static PW cutoff of 550 eV. Otherwise default VASP settings has been used, as defined for the version used. Also note that we do not use a explicit stress cutoff, but rely strictly on the maximum force cutoff.

We would like to stress the fact that the statically chosen PW cutoff and k-point grids are purely an artifact of needing higher cutoffs for the calculation of the H$_2$ dissociation curve and at the same time focusing on simplicity and transferability of the same protocol definition across the demonstrators in this work. The purpose of this work is not to present state-of-the-art accuracy or precision. And in order to make the main content to the point, we did not want to utilize dedicated convergence workchains to enable a more tailored and production ready result. However, we hope the readers will, after reading, appreciate that adding such a workchain would be straightforward. For similar reasons, as can be seen, we use the same Gaussian smearing and a rather wide smearing width for all demonstrators.

\subsubsection{Supported calculation modes}

The \texttt{relax\_type} argument supports all defined values in this work, while the \texttt{spin\_type} argument supports the values \texttt{`none'} and \texttt{`collinear'}. In case of \texttt{`collinear'} calculations, \texttt{magnetization\_per\_site} can be supplied to indicate initial magnetic moment per site in units of Bohr magneton. In the case this is not provided, the default in VASP is used, which is one Bohr magneton per atom.
The keyword \texttt{threshold\_stress} is ignored for the reason explained in the previous paragraph. Finally, the \texttt{reference\_workchain} argument guarantees that the k-points mesh that is used is identical to that used in the reference workchain. The \texttt{electronic\_type} is not used for VASP and we used Gaussian smearing for all calculations as described in the previous section.

\subsubsection{Running in the Quantum Mobile virtual machine}

Due to licensing \vasp{} is not provided in the Quantum Mobile. However, it is straightforward to compile \vasp{} and add it to Quantum Mobile for licence holders. The AiiDA-\vasp{} plugin is however included in Quantum Mobile as part of this work.

\section{Instructions to run the test cases in the Quantum Mobile.}
We present here the full set of instructions that are needed to reproduce the results presented in the manuscript, using the Quantum Mobile.
N.B.: since the results reported in the paper are obtained with very stringent choices of parameters (\texttt{`precise'} protocol), the commands listed below might start computationally-demanding simulations.
In addition, the quantum-engine simulations are run with two processors by default.
For Quantum Mobile v21.05.1 the included compiled binaries for \abinit{}, \cptwok{}, \fleur{} and \siesta{} can fail for certain runs if run with more than one processor and instead have to be run in serial.
To specify the number of processors that should be used, the \gls{cli} provides the option "\texttt{-n}".
The option accepts a value for each engine of the calculation, meaning two integers must be provided for \fleur{} ("\texttt{-n 1 2}" for instance) and one for any other code ("\texttt{-n 4}").
Moreover calculations on molecules (H$_2$ and ammonia) performed with codes designed for extended systems might result in calculations that require a lot of memory.

The following list of instructions shows how the results of this paper can be reproduced.
The label \texttt{<code>} must be substituted with the name of one of the available quantum engines.
\begin{enumerate}[leftmargin=*]
    \item \href{https://quantum-mobile.readthedocs.io/en/latest/releases/versions/21.05.1.html}{Download Quantum Mobile v21.05.1} and follow the instructions for its installation. The installation just requires the Virtual Box software.
    \item Start the virtual machine, open a terminal and type "workon aiida" to activate the virtual environment where all the python packages are installed.
    \item Make sure to perform the preliminary steps required by the quantum engine of your interest, they are listed in the previous section under {\itshape Running in the Quantum Mobile virtual machine} subsection for each code.
    \item For the ammonia test case, run the command:
    \begin{lstlisting}[
    label=code:amm1,
    ]
   aiida-common-workflows launch relax -S NH3-pyramidal -p precise -- <code>
   \end{lstlisting}
   wait the end of the execution and then run:
   \begin{lstlisting}[
    label=code:amm2,
    ]
   aiida-common-workflows launch relax -S NH3-planar -p precise -- <code>
   \end{lstlisting}
   Both commands will return an output link called \texttt{total\_energy}. Run:
   \begin{lstlisting}[
    label=code:amm3,
    ]
    verdi node attributes <pk>
   \end{lstlisting}
   to explore the value of the total energy for both calculation, \texttt{<pk>} is the node pk of \texttt{total\_energy}. The difference in energy between the planar and pyramidal energy is the value reported in Figure 4 of the main text.

\item The calculation of the \gls{eos}
    for elements El =[Si, Al] is started with the command:
    \begin{lstlisting}[
    label=code:eos1,
    ]
   aiida-common-workflows launch eos -S El -p precise -- <code>
   \end{lstlisting}
   This command launches a relaxation of the atomic coordinates for a spin-less system with \texttt{`metal'} setting. All these are defaults for the command line interface implemented in the package.
   Once the calculation is over, results are obtained running:
   \begin{lstlisting}[
    label=code:eos2,
    ]
   aiida-common-workflows launch plot-eos -- <pk>
   \end{lstlisting}
   where \texttt{<pk>} is the pk of the \gls{eos} workflow, reported at run time. The command directly plots the energy versus volume data. In case a print of the values is needed, the option  "\texttt{-t -p 4 5}" can be added. The  "\texttt{-p 4 5}" indicates that the volumes and energies should be printed with 4 and 5 decimal figures respectively.

\item The calculation of the \gls{eos}
    for GeTe is started with the command:
    \begin{lstlisting}[
    label=code:eos3,
    ]
   aiida-common-workflows launch eos -S GeTe -p precise -r positions_shape -- <code>
   \end{lstlisting}
   Once the calculation is finished, the results can be obtained with the same command explained in the previous point.

\item The calculation of the \gls{eos}
    for Fe in the ferromagnetic arrangement is started with:
    \begin{lstlisting}[
    label=code:eos4,
    ]
   aiida-common-workflows launch eos -S Fe -p precise -s collinear -- <code>
   \end{lstlisting}
   The initial magnetization used is different for each \texttt{<code>} since the default is used.
   The command for obtaining the results is always the same except that a further integer can be accepted by the  "\texttt{-p}" option to indicate the decimal figures for the total magnetization ("\texttt{-p 4 5 4}") for instance.
   The calculation for the anti-ferromagnetic case is done with:
   \begin{lstlisting}[
    label=code:eos5,
    ]
   aiida-common-workflows launch eos -S Fe -p precise -s collinear --magnetization_per_site -4 4 -- <code>
   \end{lstlisting}
    The only difference respect to the ferromagnetic case is the explicit choice of an initial magnetization.

\item The H$_2$ dissociation curve is obtained with:
\begin{lstlisting}[
    label=code:dc1,
    ]
   aiida-common-workflows launch dissociation-curve -S H2 -p precise -s collinear --magnetization-per-site -1 1 -- <code>
   \end{lstlisting}
   and the results are obtained with:
   \begin{lstlisting}[
    label=code:dc2,
    ]
   aiida-common-workflows launch plot-dissociation-curve -- <pk>
   \end{lstlisting}
    The same options of the \texttt{plot-eos} command can be used to custom the results analysis.
\end{enumerate}

For simplicity, we illustrated how to run the test cases using the command line interface of the package.
Another possibility is to create a submission script for each case.
Detailed documentation on the topic can be found in the \href{https://aiida-common-workflows.readthedocs.io/en/latest/}{\texttt{aiida-common-workflow} online documentation}.


\section*{Bibliography}
\bibliographystyle{custombibstylewithdoishort.bst}
\bibliography{template}

\begin{thebibliography}{10}

\bibitem{Burke:2012}
{K.~Burke}, \emph{Perspective on density functional theory},
  \href{https://doi.org/10.1063/1.4704546}{The Journal of Chemical Physics
  }\href{https://doi.org/10.1063/1.4704546}{\textbf{136}, 150901} (2012).

\bibitem{Jones:2015}
{R.~Jones}, \emph{Density functional theory: Its origins, rise to prominence,
  and future}, \href{https://doi.org/10.1103/revmodphys.87.897}{Reviews of
  Modern Physics }\href{https://doi.org/10.1103/revmodphys.87.897}{\textbf{87},
  897} (2015).

\bibitem{Lejaeghere:2016}
{K.~Lejaeghere}, {G.~Bihlmayer}, {T.~Bjorkman} \emph{et~al.},
  \emph{Reproducibility in density functional theory calculations of solids},
  \href{https://doi.org/10.1126/science.aad3000}{Science
  }\href{https://doi.org/10.1126/science.aad3000}{\textbf{351}, aad3000}
  (2016).

\bibitem{HjorthLarsen:2017}
{A.~H. Larsen}, {J.~J. Mortensen}, {J.~Blomqvist} \emph{et~al.}, \emph{The
  atomic simulation environment{\textemdash}a Python library for working with
  atoms}, \href{https://doi.org/10.1088/1361-648x/aa680e}{Journal of Physics:
  Condensed Matter
  }\href{https://doi.org/10.1088/1361-648x/aa680e}{\textbf{29}, 273002} (2017).

\bibitem{Gjerding:2021}
{M.~Gjerding}, {T.~Skovhus}, {A.~Rasmussen} \emph{et~al.}, \emph{Atomic
  Simulation Recipes -- a Python framework and library for automated workflows}
  (2021).

\bibitem{Huber:2020}
{S.~P. Huber}, {S.~Zoupanos}, {M.~Uhrin} \emph{et~al.}, \emph{{AiiDA} 1.0, a
  scalable computational infrastructure for automated reproducible workflows
  and data provenance},
  \href{https://doi.org/10.1038/s41597-020-00638-4}{Scientific Data
  }\href{https://doi.org/10.1038/s41597-020-00638-4}{\textbf{7}} (2020).

\bibitem{Gonze:2016}
{X.~Gonze}, {F.~Jollet}, {F.~A. Araujo} \emph{et~al.}, \emph{Recent
  developments in the {ABINIT} software package},
  \href{https://doi.org/10.1016/j.cpc.2016.04.003}{Computer Physics
  Communications
  }\href{https://doi.org/10.1016/j.cpc.2016.04.003}{\textbf{205}, 106} (2016).

\bibitem{Gonze:2020}
{X.~Gonze}, {B.~Amadon}, {G.~Antonius} \emph{et~al.}, \emph{The Abinitproject:
  Impact, environment and recent developments},
  \href{https://doi.org/10.1016/j.cpc.2019.107042}{Computer Physics
  Communications
  }\href{https://doi.org/10.1016/j.cpc.2019.107042}{\textbf{248}, 107042}
  (2020).

\bibitem{Romero:2020}
{A.~H. Romero}, {D.~C. Allan}, {B.~Amadon} \emph{et~al.}, \emph{{ABINIT}:
  Overview and focus on selected capabilities},
  \href{https://doi.org/10.1063/1.5144261}{The Journal of Chemical Physics
  }\href{https://doi.org/10.1063/1.5144261}{\textbf{152}, 124102} (2020).

\bibitem{Ratcliff:2020}
{L.~E. Ratcliff}, {W.~Dawson}, {G.~Fisicaro} \emph{et~al.}, \emph{Flexibilities
  of wavelets as a computational basis set for large-scale electronic structure
  calculations}, \href{https://doi.org/10.1063/5.0004792}{The Journal of
  Chemical Physics }\href{https://doi.org/10.1063/5.0004792}{\textbf{152},
  194110} (2020).

\bibitem{Clark:2005}
{S.~J. Clark}, {M.~D. Segall}, {C.~J. Pickard} \emph{et~al.}, \emph{First
  principles methods using {CASTEP}},
  \href{https://doi.org/10.1524/zkri.220.5.567.65075}{Zeitschrift für
  Kristallographie - Crystalline Materials
  }\href{https://doi.org/10.1524/zkri.220.5.567.65075}{\textbf{220}} (2005).

\bibitem{Hutter:2013}
{J.~Hutter}, {M.~Iannuzzi}, {F.~Schiffmann} and {J.~VandeVondele}, \emph{cp2k:
  atomistic simulations of condensed matter systems},
  \href{https://doi.org/10.1002/wcms.1159}{Wiley Interdisciplinary Reviews:
  Computational Molecular Science
  }\href{https://doi.org/10.1002/wcms.1159}{\textbf{4}, 15} (2013).

\bibitem{Kuehne:2020}
{T.~D. Kühne}, {M.~Iannuzzi}, {M.~D. Ben} \emph{et~al.}, \emph{{CP}2K: An
  electronic structure and molecular dynamics software package - Quickstep:
  Efficient and accurate electronic structure calculations},
  \href{https://doi.org/10.1063/5.0007045}{The Journal of Chemical Physics
  }\href{https://doi.org/10.1063/5.0007045}{\textbf{152}, 194103} (2020).

\bibitem{fleur}
{{https://www.flapw.de}}, \emph{{T}he {J}ülich {FLAPW} code family}.

\bibitem{Frisch:2016}
{M.~J. Frisch}, {G.~W. Trucks}, {H.~B. Schlegel} \emph{et~al.}, \emph{Gaussian
  09 {Revision} {D}.01} (2016).

\bibitem{Apra:2020}
{E.~Apr{\`{a}}}, {E.~J. Bylaska}, {W.~A. de~Jong} \emph{et~al.},
  \emph{{NWChem}: Past, present, and future},
  \href{https://doi.org/10.1063/5.0004997}{The Journal of Chemical Physics
  }\href{https://doi.org/10.1063/5.0004997}{\textbf{152}, 184102} (2020).

\bibitem{Neese:2012}
{F.~Neese}, \emph{The {ORCA} program system},
  \href{https://doi.org/10.1002/wcms.81}{{WIREs} Computational Molecular
  Science }\href{https://doi.org/10.1002/wcms.81}{\textbf{2}, 73} (2011).

\bibitem{Neese:2018}
{F.~Neese}, \emph{Software update: the {ORCA} program system, version 4.0},
  \href{https://doi.org/10.1002/wcms.1327}{{WIREs} Computational Molecular
  Science }\href{https://doi.org/10.1002/wcms.1327}{\textbf{8}} (2017).

\bibitem{Giannozzi:2009}
{P.~Giannozzi}, {S.~Baroni}, {N.~Bonini} \emph{et~al.}, \emph{{QUANTUM}
  {ESPRESSO}: a modular and open-source software project for quantum
  simulations of materials},
  \href{https://doi.org/10.1088/0953-8984/21/39/395502}{Journal of Physics:
  Condensed Matter
  }\href{https://doi.org/10.1088/0953-8984/21/39/395502}{\textbf{21}, 395502}
  (2009).

\bibitem{Giannozzi:2017}
{P.~Giannozzi}, {O.~Andreussi}, {T.~Brumme} \emph{et~al.}, \emph{Advanced
  capabilities for materials modelling with Quantum {ESPRESSO}},
  \href{https://doi.org/10.1088/1361-648x/aa8f79}{Journal of Physics: Condensed
  Matter }\href{https://doi.org/10.1088/1361-648x/aa8f79}{\textbf{29}, 465901}
  (2017).

\bibitem{Soler:2002}
{J.~M. Soler}, {E.~Artacho}, {J.~D. Gale} \emph{et~al.}, \emph{The {SIESTA}
  method forab initioorder-Nmaterials simulation},
  \href{https://doi.org/10.1088/0953-8984/14/11/302}{Journal of Physics:
  Condensed Matter
  }\href{https://doi.org/10.1088/0953-8984/14/11/302}{\textbf{14}, 2745}
  (2002).

\bibitem{Garcia:2020}
{A.~Garc{\'{\i}}a}, {N.~Papior}, {A.~Akhtar} \emph{et~al.}, \emph{Siesta:
  Recent developments and applications},
  \href{https://doi.org/10.1063/5.0005077}{The Journal of Chemical Physics
  }\href{https://doi.org/10.1063/5.0005077}{\textbf{152}, 204108} (2020).

\bibitem{Kresse:1996}
{G.~Kresse} and {J.~Furthmüller}, \emph{Efficient iterative schemes forab
  initiototal-energy calculations using a plane-wave basis set},
  \href{https://doi.org/10.1103/physrevb.54.11169}{Physical Review B
  }\href{https://doi.org/10.1103/physrevb.54.11169}{\textbf{54}, 11169} (1996).

\bibitem{Kresse:1999}
{G.~Kresse} and {D.~Joubert}, \emph{From ultrasoft pseudopotentials to the
  projector augmented-wave method},
  \href{https://doi.org/10.1103/physrevb.59.1758}{Physical Review B
  }\href{https://doi.org/10.1103/physrevb.59.1758}{\textbf{59}, 1758} (1999).

\bibitem{Talirz:2020}
{L.~Talirz}, {S.~Kumbhar}, {E.~Passaro} \emph{et~al.}, \emph{Materials Cloud, a
  platform for open computational science},
  \href{https://doi.org/10.1038/s41597-020-00637-5}{Scientific Data
  }\href{https://doi.org/10.1038/s41597-020-00637-5}{\textbf{7}} (2020).

\bibitem{Stoddart:2016}
{C.~Stoddart}, \emph{Is there a reproducibility crisis in science?},
  \href{https://doi.org/10.1038/d41586-019-00067-3}{Nature } (2016).

\bibitem{Wilkinson:2016}
{M.~D. Wilkinson}, {M.~Dumontier}, {I.~J. Aalbersberg} \emph{et~al.}, \emph{The
  {FAIR} Guiding Principles for scientific data management and stewardship},
  \href{https://doi.org/10.1038/sdata.2016.18}{Scientific Data
  }\href{https://doi.org/10.1038/sdata.2016.18}{\textbf{3}} (2016).

\bibitem{Goble:2020}
{C.~Goble}, {S.~Cohen-Boulakia}, {S.~Soiland-Reyes} \emph{et~al.}, \emph{{FAIR}
  Computational Workflows}, \href{https://doi.org/10.1162/dint_a_00033}{Data
  Intelligence }\href{https://doi.org/10.1162/dint_a_00033}{\textbf{2}, 108}
  (2020).

\bibitem{Uhrin:2021}
{M.~Uhrin}, {S.~P. Huber}, {J.~Yu} \emph{et~al.}, \emph{Workflows in {AiiDA}:
  Engineering a high-throughput, event-based engine for robust and modular
  computational workflows},
  \href{https://doi.org/10.1016/j.commatsci.2020.110086}{Computational
  Materials Science
  }\href{https://doi.org/10.1016/j.commatsci.2020.110086}{\textbf{187}, 110086}
  (2021).

\bibitem{Gresch:2018}
{D.~Gresch}, {Q.~Wu}, {G.~W. Winkler} \emph{et~al.}, \emph{Automated
  construction of symmetrized Wannier-like tight-binding models from ab initio
  calculations},
  \href{https://doi.org/10.1103/physrevmaterials.2.103805}{Physical Review
  Materials
  }\href{https://doi.org/10.1103/physrevmaterials.2.103805}{\textbf{2}} (2018).

\bibitem{Pizzi:2016}
{G.~Pizzi}, {A.~Cepellotti}, {R.~Sabatini} \emph{et~al.}, \emph{{AiiDA}:
  automated interactive infrastructure and database for computational science},
  \href{https://doi.org/10.1016/j.commatsci.2015.09.013}{Computational
  Materials Science
  }\href{https://doi.org/10.1016/j.commatsci.2015.09.013}{\textbf{111}, 218}
  (2016).

\bibitem{Perdew:1996}
{J.~P. Perdew}, {K.~Burke} and {M.~Ernzerhof}, \emph{Generalized Gradient
  Approximation Made Simple},
  \href{https://doi.org/10.1103/physrevlett.77.3865}{Physical Review Letters
  }\href{https://doi.org/10.1103/physrevlett.77.3865}{\textbf{77}, 3865}
  (1996).

\bibitem{Ghosh:2000}
{D.~C. Ghosh}, {J.~Jana} and {R.~Biswas}, \emph{Quantum chemical study of the
  umbrella inversion of the ammonia molecule},
  \href{https://doi.org/10.1002/1097-461x(2000)80:1<1::aid-qua1>3.0.co;2-d}{International
  Journal of Quantum Chemistry
  }\href{https://doi.org/10.1002/1097-461x(2000)80:1<1::aid-qua1>3.0.co;2-d}{\textbf{80},
  1} (2000).

\bibitem{Cerioni:2012}
{A.~Cerioni}, {L.~Genovese}, {A.~Mirone} and {V.~A. Sole}, \emph{Efficient and
  accurate solver of the three-dimensional screened and unscreened
  Poisson{\textquotesingle}s equation with generic boundary conditions},
  \href{https://doi.org/10.1063/1.4755349}{The Journal of Chemical Physics
  }\href{https://doi.org/10.1063/1.4755349}{\textbf{137}, 134108} (2012).

\bibitem{Castro:2003}
{A.~Castro}, {A.~Rubio} and {M.~J. Stott}, \emph{Solution of
  Poisson{\textquotesingle}s equation for finite systems using plane-wave
  methods}, \href{https://doi.org/10.1139/p03-078}{Canadian Journal of Physics
  }\href{https://doi.org/10.1139/p03-078}{\textbf{81}, 1151} (2003).

\bibitem{Bengtsson:1999}
{L.~Bengtsson}, \emph{Dipole correction for surface supercell calculations},
  \href{https://doi.org/10.1103/physrevb.59.12301}{Physical Review B
  }\href{https://doi.org/10.1103/physrevb.59.12301}{\textbf{59}, 12301} (1999).

\bibitem{Makov:1995}
{G.~Makov} and {M.~C. Payne}, \emph{Periodic boundary conditions inab
  initiocalculations}, \href{https://doi.org/10.1103/physrevb.51.4014}{Physical
  Review B }\href{https://doi.org/10.1103/physrevb.51.4014}{\textbf{51}, 4014}
  (1995).

\bibitem{Goldak:1966}
{J.~Goldak}, {C.~S. Barrett}, {D.~Innes} and {W.~Youdelis}, \emph{Structure of
  Alpha {GeTe}}, \href{https://doi.org/10.1063/1.1727231}{The Journal of
  Chemical Physics }\href{https://doi.org/10.1063/1.1727231}{\textbf{44}, 3323}
  (1966).

\bibitem{Cohen:2008}
{A.~J. Cohen}, {P.~Mori-Sanchez} and {W.~Yang}, \emph{Insights into Current
  Limitations of Density Functional Theory},
  \href{https://doi.org/10.1126/science.1158722}{Science
  }\href{https://doi.org/10.1126/science.1158722}{\textbf{321}, 792} (2008).

\bibitem{web:Huber:MCA:2021}
{S.~P. Huber}, {E.~Bosoni}, {M.~Bercx} \emph{et~al.}, \emph{Common workflows
  for computing material properties using different quantum engines},
  \href{https://doi.org/https://doi.org/10.24435/materialscloud:nz-01}{Materials
  Cloud Archive } (2021), doi:10.24435/materialscloud:nz-01.

\bibitem{Broeder:2019}
{J.~Br\"oder}, {D.~Wortmann} and {S.~Bl\"ugel}, \emph{{U}sing the
  {A}ii{DA}-{FLEUR} package for all-electron abinitio electronic structure data
  generation and processing in materials science}, in \emph{In Extreme Data
  Workshop 2018 Proceedings}, \emph{IAS Series}, vol.~40, pp. p 43--48,
  Forschungszentrum J\"ulich, J\"ulich (2019).

\bibitem{Lejaeghere:2014}
{K.~Lejaeghere}, {V.~V. Speybroeck}, {G.~V. Oost} and {S.~Cottenier},
  \emph{Error Estimates for Solid-State Density-Functional Theory Predictions:
  An Overview by Means of the Ground-State Elemental Crystals},
  \href{https://doi.org/10.1080/10408436.2013.772503}{Critical Reviews in Solid
  State and Materials Sciences
  }\href{https://doi.org/10.1080/10408436.2013.772503}{\textbf{39}, 1} (2013).

\bibitem{Liu:1989}
{D.~C. Liu} and {J.~Nocedal}, \emph{On the limited memory {BFGS} method for
  large scale optimization},
  \href{https://doi.org/10.1007/bf01589116}{Mathematical Programming
  }\href{https://doi.org/10.1007/bf01589116}{\textbf{45}, 503} (1989).

\bibitem{Bitzek:2006}
{E.~Bitzek}, {P.~Koskinen}, {F.~Gähler} \emph{et~al.}, \emph{Structural
  Relaxation Made Simple},
  \href{https://doi.org/10.1103/physrevlett.97.170201}{Physical Review Letters
  }\href{https://doi.org/10.1103/physrevlett.97.170201}{\textbf{97}} (2006).

\bibitem{Setten:2018}
{M.~van Setten}, {M.~Giantomassi}, {E.~Bousquet} \emph{et~al.}, \emph{The
  {PseudoDojo}: Training and grading a 85 element optimized norm-conserving
  pseudopotential table},
  \href{https://doi.org/10.1016/j.cpc.2018.01.012}{Computer Physics
  Communications
  }\href{https://doi.org/10.1016/j.cpc.2018.01.012}{\textbf{226}, 39} (2018).

\bibitem{Torrent:2008}
{M.~Torrent}, {F.~Jollet}, {F.~Bottin} \emph{et~al.}, \emph{Implementation of
  the projector augmented-wave method in the {ABINIT} code: Application to the
  study of iron under pressure},
  \href{https://doi.org/10.1016/j.commatsci.2007.07.020}{Computational
  Materials Science
  }\href{https://doi.org/10.1016/j.commatsci.2007.07.020}{\textbf{42}, 337}
  (2008).

\bibitem{Saha:2017}
{S.~Saha}, \emph{Soft and accurate norm conserving pseudopotentials and their
  application for structure prediction}, Ph.D. thesis, University of Basel
  (2017).

\bibitem{Vandevondele:2005}
{J.~VandeVondele}, {M.~Krack}, {F.~Mohamed} \emph{et~al.}, \emph{Quickstep:
  Fast and accurate density functional calculations using a mixed Gaussian and
  plane waves approach},
  \href{https://doi.org/10.1016/j.cpc.2004.12.014}{Computer Physics
  Communications
  }\href{https://doi.org/10.1016/j.cpc.2004.12.014}{\textbf{167}, 103} (2005).

\bibitem{Weigend:2005}
{F.~Weigend} and {R.~Ahlrichs}, \emph{Balanced basis sets of split valence,
  triple zeta valence and quadruple zeta valence quality for H to Rn: Design
  and assessment of accuracy}, \href{https://doi.org/10.1039/b508541a}{Physical
  Chemistry Chemical Physics
  }\href{https://doi.org/10.1039/b508541a}{\textbf{7}, 3297} (2005).

\bibitem{Weigend:2003}
{F.~Weigend}, {F.~Furche} and {R.~Ahlrichs}, \emph{Gaussian basis sets of
  quadruple zeta valence quality for atoms H{\textendash}Kr},
  \href{https://doi.org/10.1063/1.1627293}{The Journal of Chemical Physics
  }\href{https://doi.org/10.1063/1.1627293}{\textbf{119}, 12753} (2003).

\bibitem{Prandini:2018}
{G.~Prandini}, {A.~Marrazzo}, {I.~E. Castelli} \emph{et~al.}, \emph{Precision
  and efficiency in solid-state pseudopotential calculations},
  \href{https://doi.org/10.1038/s41524-018-0127-2}{npj Computational Materials
  }\href{https://doi.org/10.1038/s41524-018-0127-2}{\textbf{4}} (2018).

\bibitem{Marzari:1999}
{N.~Marzari}, {D.~Vanderbilt}, {A.~D. Vita} and {M.~C. Payne}, \emph{Thermal
  Contraction and Disordering of the Al(110) Surface},
  \href{https://doi.org/10.1103/physrevlett.82.3296}{Physical Review Letters
  }\href{https://doi.org/10.1103/physrevlett.82.3296}{\textbf{82}, 3296}
  (1999).

\end{thebibliography}

\end{document}